\documentclass[lettersize,journal]{IEEEtran}
\usepackage{amsmath,amsfonts}
\usepackage{amsthm}
\usepackage{amssymb}
\usepackage{pifont}
\usepackage{utfsym}
\usepackage{easyReview}
\usepackage{algorithmic}
\usepackage{algorithm}
\usepackage{array}
\usepackage[caption=false,font=normalsize,labelfont=sf,textfont=sf]{subfig}
\usepackage{easyReview}
\usepackage{textcomp}
\usepackage{stfloats}
\usepackage{url}
\usepackage{bm}
\usepackage{makecell}
\usepackage{verbatim}
\usepackage{graphicx}
\usepackage{cite}
\usepackage{caption}
\usepackage{booktabs}
\makeatletter
\renewcommand{\maketag@@@}[1]{\hbox{\m@th\normalsize\normalfont#1}}%
\makeatother
\renewcommand\arraystretch{1.2}
\allowdisplaybreaks[4]

\newtheorem{Lemma}{Lemma}

\begin{document}

\title{Joint Beamforming Design for RIS-Empowered NOMA-ISAC Systems}

\author{Chunjie Wang, Xuhui Zhang, Jinke Ren, Wenchao Liu, Shuqiang Wang,~\IEEEmembership{Senior Member,~IEEE,} Yanyan Shen,\\ Kejiang Ye,~\IEEEmembership{Senior Member,~IEEE,} Chengzhong Xu,~\IEEEmembership{Fellow,~IEEE,} and Dusit Niyato~\IEEEmembership{Fellow,~IEEE}

    \thanks{Chunjie Wang is with Shenzhen Institutes of Advanced Technology, Chinese Academy of Sciences, Shenzhen 518055, China, and also with the University of Chinese Academy of Sciences, Beijing 100049, China (e-mail: cj.wang@siat.ac.cn). \textit{(Corresponding author: Yanyan Shen).}}
    
    \thanks{Xuhui Zhang and Jinke Ren are with the Future Network of Intelligence Institute, the School of Science and Engineering, and the Guangdong Provincial Key Laboratory of Future Networks of Intelligence, The Chinese University of Hong Kong, Shenzhen, Guangdong 518172, China (e-mail: xu.hui.zhang@foxmail.com; jinkeren@cuhk.edu.cn).}

    \thanks{
Wenchao Liu is with the School of Automation and Intelligent Manufacturing, Southern University of Science and Technology, Shenzhen 518055, China (e-mail: wc.liu@foxmail.com).
}
    
    \thanks{Shuqiang Wang, Yanyan Shen, and Kejiang Ye are with the Shenzhen Institutes of Advanced Technology, Chinese Academy of Sciences, Shenzhen 518055, China (e-mail: sq.wang@siat.ac.cn; yy.shen@siat.ac.cn; kj.ye@siat.ac.cn).}

    \thanks{Chengzhong Xu is with the State Key Laboratory of IOTSC, Department of Computer Science, University of Macau, Macau, SAR, China (e-mail: czxu@um.edu.mo).}

    \thanks{Dusit Niyato is with the College of Computing and Data Science, Nanyang Technological University, Singapore 639798 (e-mail: dniyato@ntu.edu.sg).}
    
}

\maketitle

\begin{abstract}
	This paper investigates a reconfigurable intelligent surface (RIS)-assisted integrated sensing and communication (ISAC) system and proposes a joint communication and sensing beamforming design based on non-orthogonal multiple access (NOMA) technology. The system employs a dual-functional base station (DFBS) to simultaneously serve multiple users and sense multiple targets with the aid of RIS. To maximize the sum-rate of users, we jointly optimize the DFBS’s active beamforming, the RIS’s reflection coefficients, and the radar receive filters. The optimization is performed under constraints including the radar signal-to-noise ratio thresholds, the user signal-to-interference-plus-noise ratio requirements, the phase shifts of the RIS, the total transmit power, the receive filters, and the successive interference cancellation decoding order. 
    To tackle the complex interdependencies and non-convex nature of the optimization problem, we introduce an effective iterative algorithm based on the alternating optimization framework. 
    Simulation results demonstrate that the proposed algorithm outperforms baseline algorithms, highlighting its distinct advantages in the considered RIS-empowered NOMA-ISAC systems.
\end{abstract}


\section{Introduction}
\IEEEPARstart {A}{fter} decades of evolution, traditional communication systems have made significant progress in information transmission capabilities.
However, the communication and sensing systems have long occupied independent frequency bands, making it difficult to efficiently reuse potential spectrum such as high-band millimeter waves due to hardware limitations \cite{9737357}.
In addition, their functional synergy capabilities are weak, making it hard to meet the integration requirements of real-time communication and environmental sensing in emerging scenarios, such as vehicle-to-everything and the low-altitude economy \cite{10551400, 10418473}. 

Integrated sensing and communication (ISAC) has become one of the promising technologies to solve the above challenges. The core advantage of ISAC technology is to achieve deep coordination of communication and sensing functions through spectrum sharing and hardware multiplexing, breaking the barrier of traditional system frequency division operation \cite{10536135, 10233771}. Its unified waveform design enables wireless signals to simultaneously perform data transmission and environmental sensing. This dual functionality not only eliminates spectrum waste but also reduces deployment costs \cite{10012421}. In \cite{10851318}, a low-complexity ISAC system model enabled by autonomous aerial vehicles was proposed, aiming to improve the spectral efficiency and information-outage probability. While in \cite{10716474}, the power consumption of the ISAC system was minimized by optimizing the transmit beamforming. This optimization ensured the quality of service and met the Cramér–Rao bound (CRB) for radar target sensing accuracy. To maximize the energy efficiency of ISAC systems, \cite{10777502} proposed an optimization problem that jointly optimized the transmit and receive beamforming, and designed artificial noise. To solve the formulated non-convex problem, an algorithm based on semi-definite relaxation, successive convex approximation (SCA), and fractional programming was proposed. \cite{10757330} integrated the orthogonal frequency division multiplexing technology into the ISAC system to achieve super-resolution distance estimation and communication.

While existing ISAC works have significantly improved the collaboration efficiency between communication and sensing systems, their simultaneous efforts of accommodating diverse sensing applications and enabling massive connectivity inevitably trigger substantial growth in network traffic loads. The integration of non-orthogonal multiple access (NOMA) technology with ISAC has demonstrated substantial potential in solving this issue \cite{9693417, 9992245}. NOMA technology overcomes the limitation of traditional orthogonal multiple access via power domain overlay transmission and successive interference cancellation (SIC) mechanism. When supporting massive connections, it can significantly improve spectral efficiency and reduce inter-user interference. Accordingly, \cite{10608079} proposed a multiple-input multiple-output NOMA-ISAC framework and analyzed its performance in both downlink and uplink scenarios. For each scenario, the optimization of performance metrics, including sensing rate, communication rate, and outage probability, was investigated. \cite{10664461} investigated the issue of secure beamforming for a NOMA-assisted ISAC system, where targets acquiring group-oriented information from the base station (BS) of the ISAC system were cooperative. To further enhance sensing efficiency, \cite{10129092} proposed a joint optimization approach for beamforming, NOMA transmission duration, and target sensing scheduling. Meanwhile, \cite{10759667} investigated a NOMA downlink scenario in ISAC systems by integrating unmanned aerial vehicle (UAV) technology. The objective was to enhance the average achievable rate by jointly optimizing the UAV trajectory and beamforming configurations at both the BS and UAV.

Reconfigurable intelligent surface (RIS), also known as intelligent reflecting surface, is the core infrastructure for building the sixth-generation mobile communication system. The typical RIS structure is composed of a two-dimensional planar array of passive reflecting elements \cite{9722893, 10012694}. By controlling the specific parameters of the electronic circuit associated with each element, the electromagnetic characteristics of the incident signal can be adjusted, such as amplitude and phase shift \cite{9326394, 9417469}. This allows RIS to flexibly optimize the wireless channel environment in a low-power manner, and effectively improve the performance of information and energy transmission \cite{10721288, 10288203} and target detection \cite{10643002, 9454375}. In ISAC systems, the introduction of RIS can enhance spatial degrees of freedom by generating additional signal paths, enabling more accurate target sensing. Simultaneously, through low-power passive signal modulation, it optimizes wireless channels to boost communication rate and reliability. \cite{10440056} and \cite{10364735} demonstrated the advantages of integrating RIS into ISAC systems to enhance radar sensing capabilities. \cite{9769997} also investigated the potential of employing RIS in ISAC systems for improving both radar sensing and communication functionalities. The objective of maximizing the radar output signal-to-interference-plus-noise ratio (SINR) was achieved by jointly designing the BS transmit waveform and the passive beamforming of RIS. In \cite{10005150}, the energy efficiency maximization problem was studied in a RIS-empowered ISAC system by jointly optimizing the transmit beamforming and the RIS phase shift matrix under both the perfect channel state information (CSI) and the imperfect CSI cases. To address the security concerns in ISAC systems, \cite{10143420} considered the sensing target as a potential eavesdropper and introduced a RIS to enhance the physical layer security of the ISAC system. In addition, there have been a few attempts to introduce RIS in NOMA-ISAC systems \cite{10776025, 10561466, 10423585}. Particularly, \cite{10776025} used RIS to create virtual line-of-sight links in a NOMA-aided ISAC system for multi-user communication and target sensing. Meanwhile, \cite{10561466} developed a RIS-aided NOMA-ISAC scheme that supported simultaneous service for multiple downlink users and target sensing, improving the quality of service and coverage. \cite{10423585} capitalized on the extra degrees of freedom from RIS to strengthen secure communication and enable effective target detection in NOMA-assisted ISAC systems.

\begin{table*}[t]
	\caption{Comparison of Related Works}
	\centering
	\label{table1}
	\resizebox{\textwidth}{!}{
		\renewcommand{\arraystretch}{2.4} 
		\begin{tabular}{|c|c|c|c|c|c|c|c|}
			\hline
			\textbf {Ref.} & \textbf{Type of system} & \textbf{Sensing metric} & \textbf{Multi-targets} & \textbf{\makecell{Direct \& reflected \\sensing links}} & \textbf{Receive filters} & \textbf{Matched filters} & \textbf{SIC decoding order}\\
                \hline
			\cite{10664461} & \makecell{NOMA-ISAC} & \makecell{SNR} & \checkmark & \ding{53} & \checkmark & \ding{53} & \ding{53}\\
                \hline
			\cite{10129092} & \makecell{NOMA-ISAC} & \makecell{Mutual information} & \checkmark & \ding{53} & \ding{53} & \ding{53} & \ding{53}\\
                \hline
			\cite{10759667} & \makecell{NOMA-ISAC} & \makecell{CRB} & \ding{53} & \ding{53} & \ding{53} & \ding{53} & \ding{53}\\
			\hline
			\cite{10440056} & \makecell{RIS-ISAC} & \makecell{CRB} & \ding{53} & \ding{53} & \ding{53} & \ding{53} & \ding{53}\\
			\hline
			\cite{10364735} & \makecell{RIS-ISAC} & \makecell{SNR \& CRB} & \ding{53} & \checkmark & \checkmark & \checkmark & \ding{53}\\
                \hline
			\cite{9769997} & \makecell{RIS-ISAC} & \makecell{SINR} & \ding{53} & \checkmark & \checkmark & \ding{53} & \ding{53}\\
			\hline
			\cite{10005150} & \makecell{RIS-ISAC} & \makecell{\ding{53}} & \ding{53} & \checkmark & \ding{53} & \ding{53} & \ding{53}\\
			\hline
			\cite{10143420} & \makecell{RIS-ISAC} & \makecell{SNR} & \ding{53} & \ding{53} & \ding{53} & \ding{53} & \ding{53}\\
			\hline
			\cite{10776025} & \makecell{RIS-aided NOMA-ISAC} & \makecell{SNR} & \ding{53} & \checkmark & \ding{53} & \ding{53} & \checkmark\\
			\hline
			\cite{10561466} & \makecell{RIS-aided NOMA-ISAC} & \makecell{\ding{53}} & \ding{53} & \checkmark & \ding{53} & \ding{53} & \checkmark\\
			\hline
			\cite{10423585} & \makecell{RIS-aided NOMA-ISAC} & \makecell{Sensing power} & \checkmark & \ding{53} & \ding{53} & \ding{53} & \ding{53}\\
			\hline
			\textbf{Our work} & \makecell{\textbf{RIS-aided NOMA-ISAC}} & \makecell{\textbf{SNR}} & \textbf{\ding{52}} & \textbf{\ding{52}} & \textbf{\ding{52}} & \textbf{\ding{52}} & \textbf{\ding{52}}\\
			\hline
	\end{tabular}}
    \renewcommand{\arraystretch}{2.4} 
\end{table*}

To facilitate reading, a comparison between our work and related studies is presented in Table I. Existing NOMA-ISAC research \cite{10664461, 10129092, 10759667} completely rely on the direct communication links between the BSs and users/targets, facing performance fluctuations caused by path fading and blockage, and lacking controllable spatial diversity. To address these issues, some works \cite{10440056, 10364735, 9769997, 10005150, 10143420} have introduced RIS into the ISAC systems to improve the channel through controllable reflections. However, these studies do not deal with the interference among users. When the network traffic surges or a large number of users access, the efficiency of the system will be greatly reduced. Recent efforts \cite{10776025, 10561466, 10423585} have integrated NOMA and RIS into ISAC systems. However, a notable limitation of these works is the absence of matched filters and receive filters, which can effectively suppress echo noise and substantially improve sensing performance. Additionally, \cite{10776025, 10561466} only model the single-target echoes, ignoring the mutual interference between multiple targets. \cite{10561466} does not include the sensing performance of the target, weakening the sensing function. While \cite{10423585} supports joint NOMA and RIS design, it ignores the optimization of SIC decoding order and the impact of direct links, significantly restricting the performance ceiling of the system. 

Motivated by these limitations, this paper proposes a RIS-empowered NOMA-ISAC system that synchronously optimizes the transmit beamforming, RIS reflection coefficients, and radar receive filters. The optimization is performed under multiple constraints, including radar signal-to-noise ratio (SNR) thresholds, user SINR requirements, RIS phase shift constraints, total transmit power limitations, receive filters constraints, and SIC decoding order constraints. The objective is to maximize multi-user communication performance while ensuring target detection performance.
The main contributions of this work are summarized as follows:
\begin{itemize}
	\item This paper presents a novel RIS-empowered ISAC system that integrates NOMA technology, enabling the simultaneous downlink multi-user communication and multi-target detection with the objective of maximizing the sum-rate of users. Multiple constraints are systematically addressed, including the SINR requirements of users, the SNR requirements for target sensing, the RIS phase-shift configurations, the total transmit power limitation, the receive filters design, and the SIC decoding sequences.
	
	\item To address the formulated non-convex optimization problem, we employ an alternating optimization (AO) technique. The sum-rate optimization problem is decomposed into three distinct subproblems. By leveraging first-order Taylor approximation, second-order cone programming, and various mathematically creative tactics, we reformulate each subproblem into a convex form, enabling efficient solution methods. By iteratively optimizing these subproblems in an alternating manner, the optimal solution to the original non-convex problem is gradually approached.
	
	\item Comprehensive simulation studies are conducted to validate the effectiveness of the proposed algorithm by comparing it with several baseline algorithm. The results show that the proposed algorithm can successfully achieve a higher sum-rate while simultaneously maintaining the required sensing performance for targets under diverse scenarios.
	
\end{itemize}

{\textit{Organizations:}}
The remainder of this paper is structured as follows. 
Section II introduces the RIS-empowered NOMA-ISAC system, and formulates the problem of maximizing the sum-rate of users. 
In Section III, the optimization problem is decomposed into three subproblems, and an iterative algorithm is proposed to solve them. 
Section IV presents numerical results to validate the performance superiority of the proposed algorithm through comparison with other baseline algorithms. 
Finally, Section V concludes this paper.  

\textit{Notations:}
The notation introduced in this paper is summarized below. $ {{\mathbb{C}}^{M\times N}} $ represents the $ M \times N $ complex matrix. $ \mathbb{E}\{ \cdot \} $ represents the statistical expectation. $ {\mathsf{diag}}({\bf{w}}) $ indicates a diagonal matrix where the elements along the diagonal correspond to those in vector $ {\bf{w}} $. $ \mathrm{j} $ signifies the imaginary unit, where $ {\mathrm{j}^2} = -1 $. $ {\cal C}{\cal N}(\mu ,{\sigma ^2}) $ denotes the circularly symmetric complex Gaussian distribution with a mean of $ \mu $ and variance $ {\sigma ^2} $. $ \left| w \right| $ denotes the magnitude of a scalar $ w $, and $ ||{\bf{w}}|| $ denotes the norm of a vector $ {\bf{w}} $. $ {\mathsf{vec}}\{{{\bf{A}}}\} $ vectorizes the matrix $ {\bf{A}} $. The Kronecker product between two matrices $ \bf{X} $ and $ \bf{Y} $ is denoted by $ \bf{X} \otimes \bf{Y} $. For a generic matrix $ {\bf{G}} $, $ {{\bf{G}}^{\mathsf{H}}} $, $ {{\bf{G}}^*} $, and $ {{\bf{G}}^{\mathsf{T}}} $ denote the conjugate transpose, the conjugate, and the transpose, respectively. The optimal value of an optimization variable $ x $ is represented by $ x^\star $. $ \Re \{ \cdot \} $ and $ \Im \{ \cdot \} $ denote the real and imaginary parts of a complex value, respectively.

\section{System Model and Problem Formulation}

\begin{figure}[t]
	\centering
	\includegraphics[width=1\linewidth]{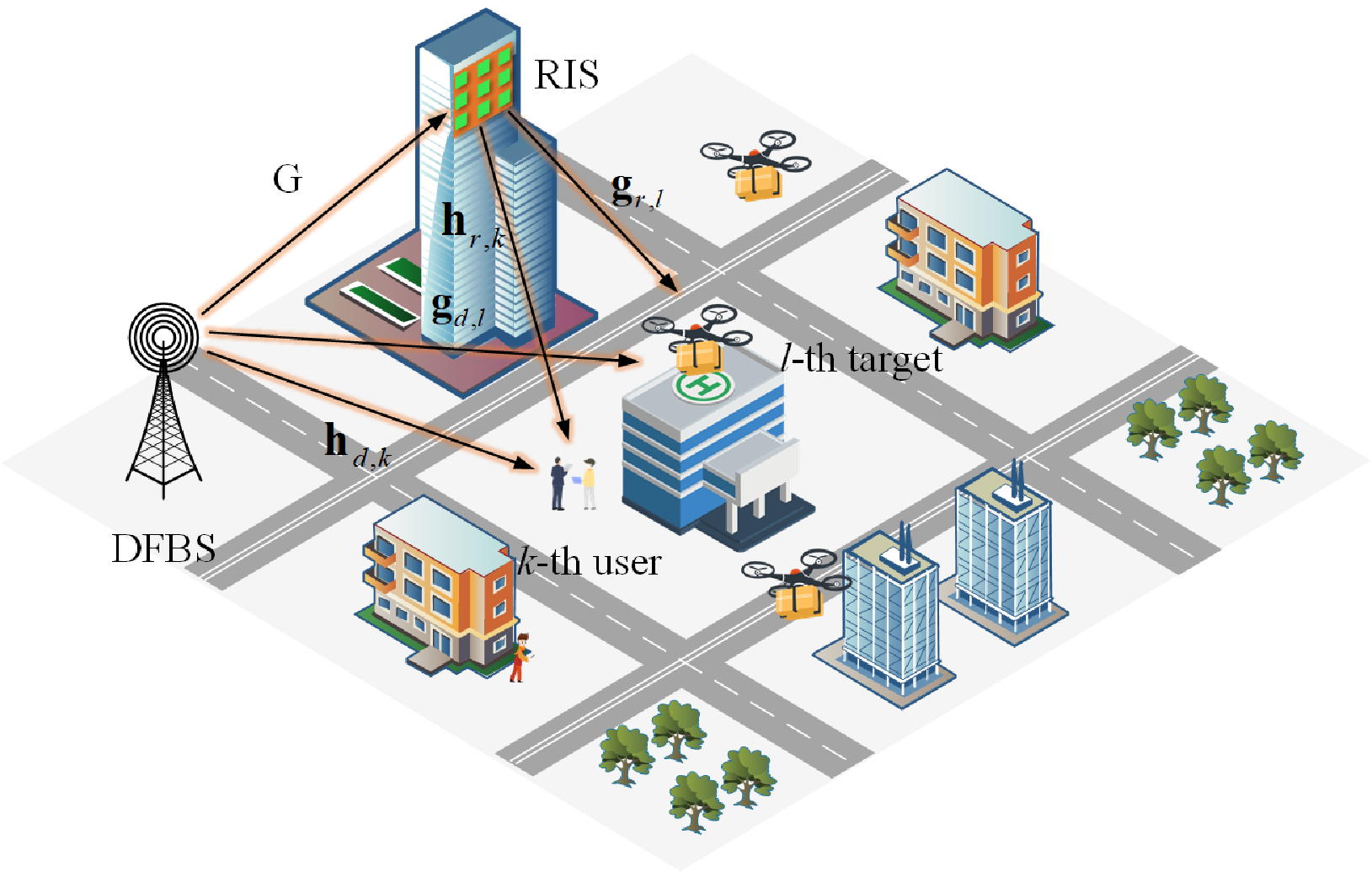}
	\captionsetup{justification=centering}
	\caption{An illustration of the RIS-empowered ISAC system model with multi-antenna DFBS and multiple users/targets.}
	\label{fig:1}
\end{figure}

As illustrated in Fig. \ref{fig:1}, we consider a RIS-empowered NOMA-ISAC system. The system core is a co-located dual-functional base station (DFBS) equipped with \(M\) transmit/receive antennas arranged in uniform linear arrays with half-wavelength spacing. The DFBS simultaneously supports two key functions: serving \(K\) single-antenna users and sensing \(L\) single-antenna targets via a RIS equipped with \(N\) reflection elements. For notational convenience, we set \({\cal N} = \{1, 2,\ldots, N\}\), \({\cal K} = \{1, 2,\ldots, K\}\), and \({\cal L} = \{1, 2,\ldots, L\}\).
Leveraging advanced self-interference mitigation techniques \cite{6832464,10364735}, the DFBS operates in full-duplex mode with perfect self-interference cancellation \cite{10052711, 10143420, 10776025, 10561466}. Regarding the radar sensing aspect, this paper focuses on a fundamental task to determine the presence of a target of interest.

To mitigate interference among targets and simplify system design, a time-division scheme
is utilized in this paper. Specifically, the total transmission duration $ T $ is divided into $ L $ time slots, which makes the number of time slots  identical to the number of targets. In each slot, the DFBS concurrently serves all $ K $ users and detects a single target. This approach effectively avoids strong cross-target interference, enabling sequential detection of $ L $ targets without impairing multi-user communication.

The transmitted signal in the $ l $-th time slot can be expressed as \cite{10472878, 10364735}
\begin{align}
	{\bf{x}}[l] = \sum\limits_{k = 1}^K {{{\bf{w}}_k}{s_{k}}[l]} + {{\bf{w}}_\vartheta}{s_\vartheta}[l] = {\bf{Ws}}[l],
\end{align}
where $ {s_{k}}[l] \in \mathbb{C} $ is the information signal for the user $ k $ with $ \mathbb{E}\{ {{{\left| {{s_{k}}[l]} \right|}^2}} \} = 1 $, and $ {{\bf{w}}_k} \in {\mathbb{C}^{M \times 1}} $ represents the corresponding beamforming vector. Then, $ {s_{\vartheta}}[l] \in \mathbb{C} $ is the sensing signal with $ \mathbb{E}\{ {{{\left| {{s_{\vartheta}}[l]} \right|}^2}} \} = 1 $ and satisfies $ \mathbb{E}\{ {{s_k}[l]s_j^{\mathsf{H}}[l]} \} = 0, j \ne k, j =  1,\ldots,K, \vartheta $, $ {{\bf{w}}_\vartheta} \in {\mathbb{C}^{M \times 1}} $ is the corresponding beamforming vector. For brevity, we define the combined beamforming matrix $ {\bf{W}} \buildrel \Delta \over = [{{\bf{w}}_1},\ldots,{{\bf{w}}_K},{{\bf{w}}_\vartheta }] $ and symbol vector $ {\bf{s}}[l] \buildrel \Delta \over = [s_1[l],\ldots,s_K[l],s_\vartheta[l]]^{\mathsf{T}} $.

\subsection{Communication Model}
Considering existing channel estimation techniques for the RIS-empowered communications \cite{10505902, 10345768}, we assume that the perfect CSI of all nodes has been obtained \cite{10561466,10364735,10472878}. Then, the received signal at the user $ k $ can be expressed as
\begin{align}
	{y_{k}}[l] = ({\bf{h}}_{d,k}^{\mathsf{H}} + {\bf{h}}_{r,k}^{\mathsf{H}}{\bf{\Phi G}}){\bf{x}}[l] + {n_k}[l],
\end{align}
where $ {\bf{h}}_{d,k} \in {\mathbb{C}^{M \times 1}} $ denotes the channel vector between the DFBS and
the user $ k $, $ {\bf{h}}_{r,k} \in {\mathbb{C}^{N \times 1}} $ represents the channel vector between the RIS and the user $ k $, and $ {\bf G} \in {\mathbb{C}^{N \times M}} $ denotes the channel matrix from the DFBS to the RIS. $ {\bf{\Phi }} \buildrel \Delta \over = \mathsf{diag}\{ {\phi _1},{\phi _2},\ldots,{\phi _N}\} \in {\mathbb{C}^{N \times N}} $ with $ {\phi _n} = {{\mathrm{e}}^{{\mathrm{j}}{\theta _n}}} $ represents the phase shift matrix of the RIS, where $ {\theta _n} \in \left[ {0,2\pi } \right) $. $ {n_k}[l] \sim {\cal C}{\cal N}(0,\sigma _k^2) $ denotes the additive white Gaussian noise (AWGN). 

To enhance the spectral efficiency, this paper implements power-domain NOMA for multi-user transmission. Following the fundamental NOMA principle, we order all users by their channel gains. It should be noted that, in the RIS-empowered systems, the NOMA implementation should jointly consider direct and reflected channel gains. Therefore, without loss of generality, the aggregated channel gains between the DFBS and users are assumed to satisfy \cite{9976948, 10670474}
\begin{align}
	0 < {\left\| {{\bf{h}}_{d,K}^{\mathsf{H}} + {\bf{h}}_{r,K}^{\mathsf{H}}{\bf{\Phi G}}} \right\|^2} \le  \ldots  \le {\left\| {{\bf{h}}_{d,1}^{\mathsf{H}} + {\bf{h}}_{r,1}^{\mathsf{H}}{\bf{\Phi G}}} \right\|^2}. \label{4}
\end{align}

The SIC decoding order is predominantly determined by channel quality. Following this principle, each user sequentially decodes and cancels signals intended for the users with poorer channel conditions (treating stronger users' signals as interference) before recovering its own information. 
In addition, in the RIS-empowered NOMA-ISAC system, radar sensing signal is embedded as artificial noise \cite{10472878, 10561466, 10776025}, which is prioritized for decoding and cancellation at all users to mitigate interference. 
Consequently, the SIC decoding sequence follows the order defined in \eqref{5}, shown at the top of the next page.
\begin{figure*}[ht]
	{\small \begin{align}
		\left\{ \begin{array}{l}
		{\left| {({\bf{h}}_{d,K}^{\mathsf{H}} + {\bf{h}}_{r,K}^{\mathsf{H}}{\bf{\Phi G}}){{\bf{w}}_\vartheta}} \right|^2} \ge {\left| {({\bf{h}}_{d,K}^{\mathsf{H}} + {\bf{h}}_{r,K}^{\mathsf{H}}{\bf{\Phi G}}){{\bf{w}}_K}} \right|^2} \ge \mathop {\max }\limits_{j = 1,\ldots,K - 1} {\left| {({\bf{h}}_{d,K}^{\mathsf{H}} + {\bf{h}}_{r,K}^{\mathsf{H}}{\bf{\Phi G}}){{\bf{w}}_j}} \right|^2},\\
		\vdots \\
		{\left| {({\bf{h}}_{d,k}^{\mathsf{H}} + {\bf{h}}_{r,k}^{\mathsf{H}}{\bf{\Phi G}}){{\bf{w}}_\vartheta}} \right|^2} \ge {\left| {({\bf{h}}_{d,k}^{\mathsf{H}} + {\bf{h}}_{r,k}^{\mathsf{H}}{\bf{\Phi G}}){{\bf{w}}_K}} \right|^2} \ge \ldots \ge {\left| {({\bf{h}}_{d,k}^{\mathsf{H}} + {\bf{h}}_{r,k}^{\mathsf{H}}{\bf{\Phi G}}){{\bf{w}}_k}} \right|^2} \ge \mathop {\max }\limits_{j = 1,\ldots,k - 1} {\left| {({\bf{h}}_{d,k}^{\mathsf{H}} + {\bf{h}}_{r,k}^{\mathsf{H}}{\bf{\Phi G}}){{\bf{w}}_j}} \right|^2},\\
		\vdots \\
		{\left| {({\bf{h}}_{d,1}^{\mathsf{H}} + {\bf{h}}_{r,1}^{\mathsf{H}}{\bf{\Phi G}}){{\bf{w}}_\vartheta}} \right|^2} \ge {\left| {({\bf{h}}_{d,1}^{\mathsf{H}} + {\bf{h}}_{r,1}^{\mathsf{H}}{\bf{\Phi G}}){{\bf{w}}_K}} \right|^2} \ge \ldots \ge {\left| {({\bf{h}}_{d,1}^{\mathsf{H}} + {\bf{h}}_{r,1}^{\mathsf{H}}{\bf{\Phi G}}){{\bf{w}}_2}} \right|^2} \ge {\left| {({\bf{h}}_{d,1}^{\mathsf{H}} + {\bf{h}}_{r,1}^{\mathsf{H}}{\bf{\Phi G}}){{\bf{w}}_1}} \right|^2},
		\end{array} \right. \label{5}
	\end{align}}
    {\noindent}\rule[0pt]{18cm}{0.04em}
\end{figure*}%
Following the above procedure, the users first demodulate and remove the radar sensing signal, and the SINR at the user $ k $ can be expressed as
\begin{align}
	{\gamma _{k \to \vartheta}} = \frac{{{{\left| {({\bf{h}}_{d,k}^{\mathsf{H}} + {\bf{h}}_{r,k}^{\mathsf{H}}{\bf{\Phi G}}){{\bf{w}}_\vartheta}} \right|}^2}}}{{\sum\limits_{i = 1}^K {{{\left| {({\bf{h}}_{d,k}^{\mathsf{H}} + {\bf{h}}_{r,k}^{\mathsf{H}}{\bf{\Phi G}}){{\bf{w}}_i}} \right|}^2} + \sigma _k^2} }}, k \in {\cal K}.
\end{align}
Following the SIC principles, each user $ k $ $ (1 \le k \le K - 1) $ sequentially decodes signals from higher-power users $ (j=k+1,\ldots, K-1, K) $, yielding the SINR for decoding the $ j $-th user’s signal at the $ k $-th user as
\begin{align}
	{\gamma _{k \to j}} = \frac{{{{\left| {({\bf{h}}_{d,k}^{\mathsf{H}} + {\bf{h}}_{r,k}^{\mathsf{H}}{\bf{\Phi G}}){{\bf{w}}_j}} \right|}^2}}}{{\sum\limits_{i = 1}^{j - 1} {{{\left| {({\bf{h}}_{d,k}^{\mathsf{H}} + {\bf{h}}_{r,k}^{\mathsf{H}}{\bf{\Phi G}}){{\bf{w}}_i}} \right|}^2} + \sigma _k^2} }},k + 1 \le j.
\end{align}
Then, the user $ k $ decodes its own signal while treating the residual lower-power users' signals as interference. Defining the user subset $ {{\cal K}_1} = \{ 2,3,\ldots,K\} $, the SINR for the user $ k \in {{\cal K}_1} $ is expressed as
\begin{align}
	{\gamma _k} = \frac{{{{\left| {({\bf{h}}_{d,k}^{\mathsf{H}} + {\bf{h}}_{r,k}^{\mathsf{H}}{\bf{\Phi G}}){{\bf{w}}_k}} \right|}^2}}}{{\sum\limits_{i = 1}^{k - 1} {{{\left| {({\bf{h}}_{d,k}^{\mathsf{H}} + {\bf{h}}_{r,k}^{\mathsf{H}}{\bf{\Phi G}}){{\bf{w}}_i}} \right|}^2} + \sigma _k^2} }},k \in {{\cal K}_1}.
\end{align}
If $ k = 1 $, the decoded SINR is
\begin{align}
	{\gamma _1} = \frac{{{{\left| {({\bf{h}}_{d,1}^{\mathsf{H}} + {\bf{h}}_{r,1}^{\mathsf{H}}{\bf{\Phi G}}){{\bf{w}}_1}} \right|}^2}}}{{\sigma _1^2}}.
\end{align}
To ensure successful interference cancellation in the NOMA systems, the target SINR for decoding $ s_k $ at the user $ k $ depends on the minimum SINR required for decoding $ s_k $ at the other users \cite{10244159, 9741332}. Therefore, we can obtain
\begin{align}
	\min \{ {\gamma _{k - 1 \to k}},{\gamma _{k - 2 \to k}},\ldots,{\gamma _{1 \to k}}\} \ge {\gamma _k},k \in {{\cal K}_1}.
\end{align}
Finally, the data rate of the user $ k $ can be expressed as
\begin{align}
	{R_k} = {\log _2}(1 + {\gamma _k}).
\end{align}

\subsection{Sensing Model}
In our system, signals from the DFBS reach targets via direct and RIS-reflected links, then reflected back along the same paths after scattering. Thus, the collected echo signal from the $ l $-th target is given by
\begin{align}
	{{\bf {y}}_{r}}[l] = \alpha _l{{\bf{G}}_l}{\bf{Ws}}[l] + {{\bf{n}}_{r}}[l],
\end{align}
where $ \alpha_l $ denotes the target radar cross section with $ {\alpha _l} \sim {\cal C}{\cal N}(0,\sigma _l^2) $, and $ {{\bf{G}}_l} \buildrel \Delta \over = ({{\bf{g}}_{d,l}} + {{\bf{G}}^{\mathsf{H}}}{\bf{\Phi }}{{\bf{g}}_{r,l}})({\bf{g}}_{d,l}^{\mathsf{H}} + {\bf{g}}_{r,l}^{\mathsf{H}}{\bf{\Phi G}}) $. $ {\bf{g}}_{d,l} \in {\mathbb{C}^{M \times 1}} $ and $ {\bf{g}}_{r,l} \in {\mathbb{C}^{N \times 1}} $ respectively denote the channel vectors between the DFBS/RIS and the target. $ {{\bf{n}}_{r}}[l] \sim {\cal C}{\cal N}({\bf{0}},\varepsilon_{l}^2{{\bf{I}}_M}) $ denotes the AWGN.
For radar sensing, the echo signals are the key to detection. Thus, by stacking $ Q $ samples\footnote{``$ Q $ samples" refer to $ Q $ time-domain echo samples that are continuously transmitted and received for radar detection of a specific target within the same time slot.
By performing cumulative processing (such as matched filtering) on these $ Q $ echo samples, the detection SNR of the target can be improved \cite{10364735,10496515}.}, we can obtain the combined received signal as \cite{10440056, 10364735}
\begin{align}
	{{\bf{Y}}_{l}} = {\alpha _l}{{\bf{G}}_l}{\bf{W}}{\bf S}_l + {{\bf{N}}_l},
\end{align}
where the symbol matrix $ {\bf{S}}_l \buildrel \Delta \over = [{\bf{s}}[l]^{(1)},{\bf{s}}[l]^{(2)},\ldots,{\bf{s}}[l]^{(Q)}] $ and the noise matrix $ {{\bf{N}}_l} \buildrel \Delta \over = [{{\bf{n}}_r}[l]^{(1)},{{\bf{n}}_r}[l]^{(2)},\ldots,{{\bf{n}}_r}[l]^{(Q)}] $.
To improve the target detection capability, a matched filter is employed to process the echo signals\footnote{``A matched filter" refers to a filter whose impulse response has a conjugate matching relationship with the transmitted signal waveform \( {\bf{S}}_l \). It performs a correlation operation (i.e., \( {{\bf{Y}}_{l}} {\bf{S}}_l^{\mathsf{H}} \)) to achieve coherent integration of the target echo. When there is a target in the received signal, the target signal has a strong correlation with \( {\bf{S}}_l \). The correlation operation can effectively superimpose the energy of the target signal with accumulation, while the noise energy cannot be enhanced synchronously due to its randomness, thus significantly improving the SNR \cite{10054402}.}. Then, \( {{\bf{Y}}_l} \) can be represented as
\begin{align}
	{\widetilde {\bf{Y}}_l} = {\alpha _l}{{\bf{G}}_l}{\bf{W}}{\bf{S}}_l{{\bf{S}}_l^{\mathsf{H}}} + {{\bf{N}}_l}{{\bf{S}}_l^{\mathsf{H}}}.
\end{align}
Let $ {\widetilde {\bf{y}}_l} \buildrel \Delta \over = {\mathsf{vec}}\{{\widetilde {\bf{Y}}_l}\} $, $ {\widetilde {\bf{w}}} \buildrel \Delta \over = {\mathsf{vec}}\{{{\bf{W}}}\} $, and $ {\widetilde {\bf{n}}_l} \buildrel \Delta \over = {\mathsf{vec}} \{{{\bf{N}}_l}{{\bf{S}}_l^{\mathsf{H}}}\} $, the echo signal can be rewritten as
\begin{align}
	{\widetilde {\bf{y}}_l} = {\alpha _l}({\bf{S}}_l{{\bf{S}}_l^{\mathsf{H}}} \otimes {{\bf{G}}_l})\widetilde {\bf{w}} + {\widetilde {\bf{n}}_l}.
\end{align}
To process the signal $ {\widetilde {\bf{y}}_l} $, we apply the receive filter ${\bf{u}}_l \in {\mathbb{C}^{M \times (K+1)}}$, leading to
\begin{align}
	{{\bf{u}}_l^{\mathsf{H}}}{\widetilde {\bf{y}}_l} = {\alpha _l}{{\bf{u}}_l^{\mathsf{H}}}({\bf{S}}{{\bf{S}}^{\mathsf{H}}} \otimes {{\bf{G}}_l})\widetilde {\bf{w}} + {{\bf{u}}_l^{\mathsf{H}}}{\widetilde {\bf{n}}_l}.
\end{align}
Consequently, the hypothesis testing problem for the radar receiver output is formulated as
\begin{align}
	\textit{z}_l = \left\{ \begin{array}{l}
		{{\cal H}_0}:{{\bf{u}}_l^{\mathsf{H}}}{\widetilde {\bf{n}}_l},\\
		{{\cal H}_1}:{\alpha _l}{{\bf{u}}_l^{\mathsf{H}}}({\bf{S}}{{\bf{S}}^{\mathsf{H}}} \otimes {{\bf{G}}_l})\widetilde {\bf{w}} + {{\bf{u}}_l^{\mathsf{H}}}{\widetilde {\bf{n}}_l},
	\end{array} \right.
\end{align}
where event $ {{\cal H}_0} $ denotes the absence of a target (null hypothesis), and event $ {{\cal H}_1} $ represents the presence of a target (alternative hypothesis). The conditional probability distributions of the test statistic $ \textit{z}_l $ under these hypotheses are given by  $ \textit{z}_l|{{\cal H}_0} \sim {\cal C}{\cal N}(0,{\beta _0}) $ and $ \textit{z}_l|{{\cal H}_1} \sim {\cal C}{\cal N}(0,{\beta _1}) $ with $ {\beta _0} = Q\varepsilon_{l}^2{{\bf{u}}_l^{\mathsf{H}}}{\bf{u}}_l $ and $ {\beta _1} = \sigma _l^2\mathbb{E}\{ {\left| {{\bf{u}}_l^{\mathsf{H}}}({\bf{S}}{{\bf{S}}^{\mathsf{H}}} \otimes {{\bf{G}}_l})\widetilde {\bf{w}} \right|^2}\} + Q\varepsilon_{l}^2{{\bf{u}}_l^{\mathsf{H}}}{\bf{u}}_l $. 

According to \cite{10364735, 9531484}, the detection probability exhibits a monotonic increase with the SNR, defined as
\begin{align}
	{\rm SNR}_l = \frac{{{\beta _1}}}{{{\beta _0}}} = \frac{\sigma _l^2\mathbb{E}\{ {\left| {{\bf{u}}_l^{\mathsf{H}}}({\bf{S}}{{\bf{S}}^{\mathsf{H}}} \otimes {{\bf{G}}_l})\widetilde {\bf{w}} \right|^2}\}}{Q\varepsilon_{l}^2{{\bf{u}}_l^{\mathsf{H}}}{\bf{u}}_l}. \label{22}
\end{align}
Maximizing SNR is critical for improving target detection performance. However, the complex expectation in the numerator poses challenges for direct optimization. To address this, we utilize the following Lemma to derive a tractable lower bound.
\begin{Lemma}\rm{(\textit{Jensen’s Inequality}): For any convex function $f(\cdot)$ and random variable $x$ with finite expectiation $\mathbb{E}\{ x\}$, the following holds:}
\begin{equation}
    \mathbb{E}\{ f(x)\} \ge f(\mathbb{E}\{ x\}).\end{equation}\label{L1JensenInequality}\end{Lemma}
\noindent By applying Lemma \ref{L1JensenInequality} with $ \mathbb{E}\{ {{\bf{SS}}^{\mathsf{H}}} \} = Q{\bf{I}}_{K + 1} $, we can obtain the following lower bound for $ {\rm{SNR}}_l $:
\begin{align}
	{\rm SNR}_l \ge \frac{{Q\sigma _l^2{{\left| {{{\bf{u}}_l^{\mathsf{H}}}({{\bf{I}}_{K + 1}} \otimes {{\bf{G}}_l})\widetilde {\bf{w}}} \right|}^2}}}{{\varepsilon _l^2{{\bf{u}}_l^{\mathsf{H}}}{\bf{u}}_l}}, l \in {\cal L}.
\end{align}

\subsection{Problem Formulation}
We aim to jointly optimize the beamforming vectors ${\bf{w}}_k$ and ${\bf{w}}_\vartheta$, reflection coefficients matrix ${\bf{\Phi}}$, and receive filters ${\bf{u}}_l$ to maximize the sum-rate of users. The corresponding optimization problem is formulated as
\begin{subequations}
	\begin{align}
		&{\rm{P1}}: \mathop {\max }\limits_{{\bf{w}}_k, {\bf{w}}_\vartheta, {\bf{\Phi }}, {\bf{u}}_l}\ \sum\limits_{k = 1}^K {{R_{k}}}, \\
		&{\mathrm{s.t.}}\ \ \frac{{Q\sigma _l^2{{\left| {{{\bf{u}}_l^{\mathsf{H}}}({{\bf{I}}_{K + 1}} \otimes {{\bf{G}}_l})\widetilde {\bf{w}}} \right|}^2}}}{{\varepsilon _l^2{{\bf{u}}_l^{\mathsf{H}}}{\bf{u}}_l}} \ge {\Gamma_l}, l \in {\cal L}, \label{24b} \\
		&\min \{ {\gamma _k}, {\gamma _{k - 1 \to k}}, {\gamma _{k - 2 \to k}},\ldots, {\gamma _{1 \to k}}\} \ge {r^k_{\text{th}}}, k \in {{\cal K}_1}, \label{24c} \\
		&\left| {{\phi _n}} \right| = 1, n \in {\cal N}, \label{24d}\\
		&{\sum\limits_{k = 1}^K {\left\| {\bf{w}}_k \right\|^2} + \left\| {\bf{w}}_\vartheta \right\|^2} \le P_{\text{th}}, \label{24e}\\
		&\left\| {\bf{u}}_l \right\|^2 = 1, l \in {\cal L}, \label{24f}\\
		&\eqref{5}, \label{24g}
	\end{align}
\end{subequations}
where $ {\Gamma_l} $ denotes the radar SNR threshold for target $ l $, $ {r^k_{\text{th}}} $ is the minimum SINR requirement for the user $ k $, and $ P_{\text{th}} $ represents the maximum transmit power of the DFBS. Constraint (\ref{24d}) governs the reflection coefficients of the RIS, (\ref{24e}) imposes the transmit power limitation, (\ref{24f}) imposes unit norm on the radar receive filter, and (\ref{24g}) dictates the SIC order.

Notably, the optimization problem is non-convex because of the interdependent optimization variables in both the objective function and constraints (\ref{24b})-(\ref{24c}), as well as the existence of the unit modulus constraints (\ref{24d}) and (\ref{24f}).

\section{Joint Active and Passive Beamforming Optimization}

\subsection{Transformation of Objective Function}
To effectively address the non-convex optimization problem P1, in the subsequent subsections, we decompose it into three subproblems. Each subproblem is reformulated into a convex form via the SCA method and solved iteratively. First, the objective function of P1 is rewritten as
\begin{align}
	\sum\limits_{k = 1}^K {{R_k}} =&\ \sum\limits_{k = 2}^K {{{\log }_2}\left(\frac{{\sum\limits_{i = 1}^k {\left| {({\bf{h}}_{d,k}^{\mathsf{H}} + {\bf{h}}_{r,k}^{\mathsf{H}}{\bf{\Phi G}}){{\bf{w}}_i}} \right|^2 + \sigma _k^2} }}{{\sum\limits_{i = 1}^{k - 1} {{{\left| {({\bf{h}}_{d,k}^{\mathsf{H}} + {\bf{h}}_{r,k}^{\mathsf{H}}{\bf{\Phi G}}){{\bf{w}}_i}} \right|}^2} + \sigma _k^2} }}\right)} \notag \\ 
	&+ {\log _2}\left(\frac{{\left| {({\bf{h}}_{d,1}^{\mathsf{H}} + {\bf{h}}_{r,1}^{\mathsf{H}}{\bf{\Phi G}}){{\bf{w}}_1}} \right|^2 + \sigma _1^2}}{{\sigma _1^2}}\right).
\end{align}
To transform the objective function into a manageable form, we introduce auxiliary variables $ {\eta _k}, {\tau _k}$, and $ {\zeta _1} $. Consequently, the following constraints can be derived
\begin{align}
	&\sum\limits_{i = 1}^k {\left| {({\bf{h}}_{d,k}^{\mathsf{H}} + {\bf{h}}_{r,k}^{\mathsf{H}}{\bf{\Phi G}}){{\bf{w}}_i}} \right|^2 + \sigma _k^2} \ge {{\text{e}}^{{\eta _k}}}, k \in {\cal K}_1, \label{26} \\
	&\sum\limits_{i = 1}^{k - 1} {{{\left| {({\bf{h}}_{d,k}^{\mathsf{H}} + {\bf{h}}_{r,k}^{\mathsf{H}}{\bf{\Phi G}}){{\bf{w}}_i}} \right|}^2} + \sigma _k^2}  \le {{\text{e}}^{{\tau _k}}}, k \in {\cal K}_1, \label{27} \\
	&\left| {({\bf{h}}_{d,1}^{\mathsf{H}} + {\bf{h}}_{r,1}^{\mathsf{H}}{\bf{\Phi G}}){{\bf{w}}_1}} \right|^2 + \sigma _1^2 \ge {{\text{e}}^{{\zeta _1}}}. \label{28} 
\end{align}
Then, the objective function satisfies
\begin{align}
	\sum\limits_{k = 1}^K {{R_k}} &\ge \sum\limits_{k = 2}^K {{{\log }_2}{{\text{e}}^{{\eta _k} - {\tau _k}}}}  + {\log _2}{{\text{e}}^{{\zeta _1}}} - {\log _2}\sigma _1^2 \notag \\
	&\ge {\log _2}{\text{e}} \cdot \left(\sum\limits_{k = 2}^K {({\eta _k} - {\tau _k})}  + {\zeta _1}\right) - {\log _2}\sigma _1^2.
\end{align}
By eliminating constants and irrelevant terms, i.e., ``$ {\log _2}{\text{e}} $" and ``$ {\log _2}\sigma _1^2 $", problem P1 can be rewritten as
\begin{subequations}
	\begin{align}
		{\rm{P2}}:&\mathop {\max }\limits_{{{\bf{w}}_k},{{\bf{w}}_\vartheta },{\bf{\Phi }},{\bf{u}},\atop {\eta _k},{\tau _k},{\zeta _1}} \;\sum\limits_{k = 2}^K {({\eta _k} - {\tau _k})} + {\zeta _1}, \\
		&\quad\ \; {\mathrm{s.t.}}\quad\ \; \eqref{24b}\text{-}\eqref{24g},\eqref{26}\text{-}\eqref{28}.
	\end{align}
\end{subequations}
Then, the objective function is convex, yet non-convex constraints persist, complicating the solution to problem P2. Thus, problem P2 is decomposed into three subproblems. By fixing other variables, we sequentially solve the subproblems related to variables $\{{\bf{w}}_k, \forall k, {\bf{w}}_\vartheta\}$, $\{{\bf{u}}_l, \forall l\}$, and $\{{\bf{\Phi }}\}$. The following subsections presents a detailed analysis of the solution.

\subsection{Active Beamforming Optimization}
Define $ {\bf{H}}_k^{\mathsf{H}} \buildrel \Delta \over = {\bf{h}}_{d,k}^{\mathsf{H}} + {\bf{h}}_{r,k}^{\mathsf{H}}{\bf{\Phi G}} $, problem P2 can be reformulated as P3, as shown at the top of the next page.
\begin{figure*}[t]
	\begin{subequations}
		{\small \begin{align}
			{\rm{P3}}:& \mathop {\max }\limits_
			{{{\bf{w}}_k},{{\bf{w}}_\vartheta },\atop {\eta _k},{\tau _k},{\zeta _1}} \;\sum\limits_{k = 2}^K {({\eta _k} - {\tau _k})}  + {\zeta _1},\\
			{\mathrm{s.t.}}\  & \sum\limits_{i = 1}^k {\left| {{\bf{H}}_k^{\mathsf{H}}{{\bf{w}}_i}} \right|^2 + } \sigma _k^2 \ge {{\text{e}}^{{\eta _k}}},k \in {\cal K}_1,\label{34b} \\
			&\sum\limits_{i = 1}^{k - 1} {{{\left| {{\bf{H}}_k^{\mathsf{H}}{{\bf{w}}_i}} \right|}^2} + \sigma _k^2}  \le {{\text{e}}^{{\tau _k}}},k \in {\cal K}_1,\label{34c} \\
			&\left| {{\bf{H}}_1^{\mathsf{H}}{{\bf{w}}_1}} \right|^2 + \sigma _1^2 \ge {{\text{e}}^{{\zeta _1}}},\label{34d} \\
			&{\left| {{{\bf{u}}_l^{\mathsf{H}}}({{\bf{I}}_{K + 1}} \otimes {{\bf{G}}_l})\widetilde {\bf{w}}} \right|^2} \ge \frac{{{\Gamma _l}\varepsilon _l^2{{\bf{u}}_l^{\mathsf{H}}}{\bf{u}}_l}}{{Q\sigma _l^2}},l \in {\cal L},\label{34e}\\
			&\min \left \{ \frac{{{{\left| {{\bf{H}}_k^{\mathsf{H}}{{\bf{w}}_k}} \right|}^2}}}{{\sum\limits_{i = 1}^{k - 1} {{{\left| {{\bf{H}}_k^{\mathsf{H}}{{\bf{w}}_i}} \right|}^2} + \sigma _k^2} }},\frac{{{{\left| {{\bf{H}}_{k - 1}^{\mathsf{H}}{{\bf{w}}_k}} \right|}^2}}}{{\sum\limits_{i = 1}^{k - 1} {{{\left| {{\bf{H}}_{k - 1}^{\mathsf{H}}{{\bf{w}}_i}} \right|}^2} + \sigma _{k - 1}^2} }},\frac{{{{\left| {{\bf{H}}_{k - 2}^{\mathsf{H}}{{\bf{w}}_k}} \right|}^2}}}{{\sum\limits_{i = 1}^{k - 1} {{{\left| {{\bf{H}}_{k - 2}^{\mathsf{H}}{{\bf{w}}_i}} \right|}^2} + \sigma _{k - 2}^2} }},\ldots,\frac{{{{\left| {{\bf{H}}_1^{\mathsf{H}}{{\bf{w}}_k}} \right|}^2}}}{{\sum\limits_{i = 1}^{k - 1} {{{\left| {{\bf{H}}_1^{\mathsf{H}}{{\bf{w}}_i}} \right|}^2} + \sigma _1^2} }}\right \}  \ge r_{\text{th}}^k,k \in {{\cal K}_1},\label{34f}\\
			&\sum\limits_{k = 1}^K {{{\left\| {{{\bf{w}}_k}} \right\|}^2}}  + {\left\| {{{\bf{w}}_\vartheta }} \right\|^2} \le {P_{\text{th}}},\label{34g}\\
			&\left\{ \begin{array}{l}
				{\left| {{\bf{H}}_k^{\mathsf{H}}{{\bf{w}}_\vartheta}} \right|^2} \ge {\left| {{\bf{H}}_k^{\mathsf{H}}{{\bf{w}}_K}} \right|^2} \ge \mathop {\max }\limits_{j = 1,\ldots,K - 1} {\left| {{\bf{H}}_k^{\mathsf{H}}{{\bf{w}}_j}} \right|^2},\\
				\vdots \\
				{\left| {{\bf{H}}_k^{\mathsf{H}}{{\bf{w}}_\vartheta}} \right|^2} \ge {\left| {{\bf{H}}_k^{\mathsf{H}}{{\bf{w}}_K}} \right|^2} \ge  \ldots  \ge {\left| {{\bf{H}}_k^{\mathsf{H}}{{\bf{w}}_k}} \right|^2} \ge \mathop {\max }\limits_{j = 1,\ldots,k - 1} {\left| {{\bf{H}}_k^{\mathsf{H}}{{\bf{w}}_j}} \right|^2},\\
				\vdots \\
				{\left| {{\bf{H}}_1^{\mathsf{H}}{{\bf{w}}_\vartheta}} \right|^2} \ge {\left| {{\bf{H}}_1^{\mathsf{H}}{{\bf{w}}_K}} \right|^2} \ge  \ldots  \ge {\left| {{\bf{H}}_1^{\mathsf{H}}{{\bf{w}}_2}} \right|^2} \ge {\left| {{\bf{H}}_1^{\mathsf{H}}{{\bf{w}}_1}} \right|^2}.
			\end{array} \right.\label{34h}
		\end{align}}
		{\noindent}\rule[-10pt]{18cm}{0.04em}
	\end{subequations}
\end{figure*}%
We employ the first-order Taylor expansion to approximate the right-hand side term of the non-convex constraint \eqref{34c} at given point $ {\widehat \tau _k} $. Thus, constraint \eqref{34c} can be rewritten as 
\begin{align}
	\sum\limits_{i = 1}^{k - 1} {{{\left| {{\bf{H}}_k^{\mathsf{H}}{{\bf{w}}_i}} \right|}^2} + \sigma _k^2} \le {{\text{e}}^{{{\widehat \tau }_k}}}(1 + {\tau _k} - {\widehat \tau _k}),k \in {\cal K}.\label{35}
\end{align}
Then, by leveraging the second-order cone constraint transformation, where $ {\xi ^2} \le \varsigma \chi (\varsigma  \ge 0,\chi  \ge 0) \Rightarrow \left\| {{{[2\xi ,\ \varsigma  - \chi ]}^{\mathsf{H}}}} \right\| \le \varsigma  + \chi $, constraint (\ref{35}) is further transformed to
{\small \begin{align}
	\left\| {{{[2{\bf{H}}_k^{\mathsf{H}}{{\bf{w}}_{k - 1}},\ldots,2{\bf{H}}_k^{\mathsf{H}}{{\bf{w}}_1},2{\sigma _k},{\Delta _1} - 1]}^{\mathsf{H}}}} \right\| \le {\Delta _{1,k}} + 1, k \in {{\cal K}_1},
\end{align}}%
where $ {\Delta _{1,k}} \buildrel \Delta \over = {{\text{e}}^{{{\widehat \tau }_k}}}(1 + {\tau _k} - {\widehat \tau _k}) $. Similarly, constraint (\ref{34g}) can be rewritten as
\begin{align}
	\left\| {{{[{{\bf{w}}_\vartheta },{{\bf{w}}_K},\ldots,{{\bf{w}}_1}]}^{\mathsf{H}}}} \right\| \le \sqrt {{P_{\text{th}}}}.
\end{align}
Next, we focus on the quadratic term $ {{{\left| {{\bf{H}}_k^{\mathsf{H}}{{\bf{w}}_i}} \right|}^2}} $. By using the first-order Taylor approximation\footnote{Since $ {{{\left| {{\bf{H}}_k^{\mathsf{H}}{{\bf{w}}_i}} \right|}^2}} $ is a quadratic form of $ {\bf{w}}_i $, the local curvature is gentle and its gradient can accurately reflect the changing trend. Thus, the first-order Taylor approximation can provide a reliable convex lower bound while ensuring monotonic convergence of iterations. In contrast, second-order method requires calculating and storing a large-scale Hessian matrix, resulting in a sharp increase in memory overhead and matrix operations. Therefore, based on the trade-off between performance and complexity, we adopt the first-order approximation.}
at the point $ {\widehat {\bf{w}}_i} $, it can be expressed as \cite{10776025,10561466}
\begin{align}
	{\left| {{\bf{H}}_k^{\mathsf{H}}{{\bf{w}}_i}} \right|^2} = 2\Re \left \{\widehat {\bf{w}}_i^{\mathsf{H}}{{\bf{H}}_k}{\bf{H}}_k^{\mathsf{H}}{{\bf{w}}_i}\right \} - \Re \left \{\widehat {\bf{w}}_i^{\mathsf{H}}{{\bf{H}}_k}{\bf{H}}_k^{\mathsf{H}}{\widehat {\bf{w}}_i} \right\}. \label{38}
\end{align}
Clearly, the quadratic form in \eqref{34b} and \eqref{34d} can be replaced by the first-order Taylor approximation, and the constraints are rewritten as
\begin{align}
	&\sum\limits_{i = 1}^k {\left(2\Re \left\{\widehat {\bf{w}}_i^{\mathsf{H}}{{\bf{H}}_k}{\bf{H}}_k^{\mathsf{H}}{{\bf{w}}_i}\right\} - \Re \left\{\widehat {\bf{w}}_i^{\mathsf{H}}{{\bf{H}}_k}{\bf{H}}_k^{\mathsf{H}}{{\widehat {\bf{w}}}_i}\right\}\right)} + \sigma _k^2\qquad\qquad\quad \notag \\
	&\qquad\qquad\qquad\qquad\qquad\qquad\qquad\quad\ \ \ \ge {{\text{e}}^{{\eta _k}}},k \in {{\cal K}_1},
\end{align}
\begin{align}
	&2\Re \left\{\widehat {\bf{w}}_1^{\mathsf{H}}{{\bf{H}}_1}{\bf{H}}_1^{\mathsf{H}}{{\bf{w}}_1}\right\} - \Re \left\{\widehat {\bf{w}}_1^{\mathsf{H}}{{\bf{H}}_1}{\bf{H}}_1^{\mathsf{H}}{\widehat {\bf{w}}_1}\right\} + \sigma _1^2 \ge {{\text{e}}^{{\zeta _1}}}.
\end{align}
In terms of the SIC decoding order constraint \eqref{34h}, it is equivalent to
\begin{align}
	\left\{ \begin{array}{l}
		{\left| {{\bf{H}}_k^{\mathsf{H}}{{\bf{w}}_\vartheta}} \right|^2} \ge {\left| {{\bf{H}}_k^{\mathsf{H}}{{\bf{w}}_K}} \right|^2},k \in {\cal K}, \\
		{\left| {{\bf{H}}_k^{\mathsf{H}}{{\bf{w}}_{i + 1}}} \right|^2} \ge {\left| {{\bf{H}}_k^{\mathsf{H}}{{\bf{w}}_i}} \right|^2},i \in \{ 1,\ldots,K - 1\}.
	\end{array} \right.\label{41}
\end{align}
Then, by applying the principle in \eqref{38} to remove the quadratic form on the left-hand side of \eqref{41}, we get 
\begin{align}
	\left\{ \begin{array}{l}
		2\Re \left\{\widehat {\bf{w}}_\vartheta ^{\mathsf{H}}{{\bf{H}}_k}{\bf{H}}_k^{\mathsf{H}}{{\bf{w}}_\vartheta }\right\} - \Re \left\{\widehat {\bf{w}}_\vartheta ^{\mathsf{H}}{{\bf{H}}_k}{\bf{H}}_k^{\mathsf{H}}{\widehat {\bf{w}}_\vartheta }\right\} \\ 
		\qquad \ge {\left| {{\bf{H}}_k^{\mathsf{H}}{{\bf{w}}_K}} \right|^2},k \in {\cal K}, \\
		2\Re \left\{\widehat {\bf{w}}_{i + 1}^{\mathsf{H}}{{\bf{H}}_k}{\bf{H}}_k^{\mathsf{H}}{{\bf{w}}_{i + 1}}\right\} - \Re \left \{\widehat {\bf{w}}_{i + 1}^{\mathsf{H}}{{\bf{H}}_k}{\bf{H}}_k^{\mathsf{H}}{\widehat {\bf{w}}_{i + 1}}\right \} \\
		\qquad\ge {\left| {{\bf{H}}_k^{\mathsf{H}}{{\bf{w}}_i}} \right|^2},i \in \{ 1,\ldots,K - 1\}.
	\end{array} \right.
\end{align}
For constraint \eqref{34f}, it can be reformulated as
\begin{align}
	&\frac{{{{\left| {{\bf{H}}_j^{\mathsf{H}}{{\bf{w}}_k}} \right|}^2}}}{{\sum\limits_{i = 1}^{k - 1} {{{\left| {{\bf{H}}_j^{\mathsf{H}}{{\bf{w}}_i}} \right|}^2} + \sigma _j^2} }} \ge r_{\text{th}}^k,\ k \in {{\cal K}_1},j = 1,\ldots,k \notag \\
	&\Rightarrow {\left| {{\bf{H}}_j^{\mathsf{H}}{{\bf{w}}_k}} \right|^2} \ge r_{\text{th}}^k\left(\sum\limits_{i = 1}^{k - 1} {{{\left| {{\bf{H}}_j^{\mathsf{H}}{{\bf{w}}_i}} \right|}^2} + \sigma _j^2}\right).\label{43}
\end{align}
Then, \eqref{43} can be similarly converted into 
\begin{align}
	&2\Re \left\{\widehat {\bf{w}}_k^{\mathsf{H}}{{\bf{H}}_j}{\bf{H}}_j^{\mathsf{H}}{{\bf{w}}_k}\right\} - \Re \left\{\widehat {\bf{w}}_k^{\mathsf{H}}{{\bf{H}}_j}{\bf{H}}_j^{\mathsf{H}}{\widehat {\bf{w}}_k}\right\} \notag \\ 
	&\ge r_{\text{th}}^k\left(\sum\limits_{i = 1}^{k - 1} {2\Re \left\{\widehat {\bf{w}}_i^{\mathsf{H}}{{\bf{H}}_j}{\bf{H}}_j^{\mathsf{H}}{{\bf{w}}_i}\right\} - \Re \left\{\widehat {\bf{w}}_i^{\mathsf{H}}{{\bf{H}}_j}{\bf{H}}_j^{\mathsf{H}}{{\widehat {\bf{w}}}_i}\right\}}  + \sigma _j^2\right), \notag \\
	&\qquad\qquad\qquad\qquad\qquad\qquad\qquad k \in {{\cal K}_1},j = 1,\ldots,k,
\end{align}
Finally, for \eqref{34e}, define $ {{\bf{{\rm A}}}_l^{\mathsf{H}}} \buildrel \Delta \over = {{\bf{u}}_l^{\mathsf{H}}}({{\bf{I}}_{K + 1}} \otimes {{\bf{G}}_l}) $ and $ {\Delta _{2,l}} \buildrel \Delta \over = {{{\Gamma _l}\varepsilon _l^2{{\bf{u}}_l^{\mathsf{H}}}{\bf{u}}_l} \mathord{\left/
		{\vphantom {{{\Gamma _l}\varepsilon _l^2{{\bf{u}}^{\mathsf{H}}}{\bf{u}}} {(Q\sigma _l^2)}}} \right.
		\kern-\nulldelimiterspace} {(Q\sigma _l^2)}} $, we can obtain
\begin{align}
	&{\left| {{\bf{{\rm A}}}_l^{\mathsf{H}}\widetilde {\bf{w}}} \right|^2} \ge {\Delta _{2,l}},l \in {\cal L}, \notag \\
	&\Rightarrow 2\Re \left\{{\widehat {\widetilde {\bf{w}}}^{\mathsf{H}}}{{\bf{{\rm A}}}_l}{\bf{{\rm A}}}_l^{\mathsf{H}}\widetilde {\bf{w}}\right\} - \Re \left\{{\widehat {\widetilde {\bf{w}}}^{\mathsf{H}}}{{\bf{{\rm A}}}_l}{\bf{{\rm A}}}_l^{\mathsf{H}}\widehat {\widetilde {\bf{w}}}\right\} \ge {\Delta _{2,l}}.
\end{align}
Following the transformations outlined above, problem P3 is transformed into a convex optimization problem, which can be efficiently solved using standard convex optimization solvers.

\subsection{Receive Filter Optimization}

When other parameters are fixed, optimizing \({\bf{u}}_l\) reduces a feasibility check problem without an explicit objective function. To address this, we update \({\bf{u}}_l\) by maximizing the SNR lower bound, leading to the following optimization problem
\begin{subequations}
	\begin{align}
		{\rm{P4}}: \mathop {\max }\limits_{{\bf{u}}_l} \ &\frac{{Q\sigma _l^2{{\left| {{{\bf{u}}_l^{\mathsf{H}}}({{\bf{I}}_{K + 1}} \otimes {{\bf{G}}_l})\widetilde {\bf{w}}} \right|}^2}}}{{\varepsilon _l^2{{\bf{u}}_l^{\mathsf{H}}}{\bf{u}}_l}},\\
		{\mathrm{s.t.}}\ &{\left\| {\bf{u}}_l \right\|^2} = 1, l \in {\cal L},
	\end{align}
\end{subequations}
which is a generalized Rayleigh quotient, whose optimal solution is \cite{10054402, 10298597, 9968163}
\begin{align}
	{{\bf{u}}_l^{\star}} = \frac{{({{\bf{I}}_{K + 1}} \otimes {{\bf{G}}_l})\widetilde {\bf{w}}}}{{{{\widetilde {\bf{w}}}^{\mathsf{H}}}({{\bf{I}}_{K + 1}} \otimes {\bf{G}}_l^{\mathsf{H}}{{\bf{G}}_l})\widetilde {\bf{w}}}}, l \in {\cal L}. \label{44}
\end{align}
It is evident that \( {{\bf{u}}_l^{\star}} \) remains feasible for the original problem if the original problem is feasible. The reason is that, the solution from the previous iteration already satisfies the radar SNR constraint, and maximizing this SNR can further guarantee compliance.
		
\subsection{Passive Beamforming Optimization}
After obtaining $ {{\bf{u}}_l^{\star}} $, it can be deduced that $ e^{\jmath \widetilde \vartheta} \mathbf{u}_l^{\star} $ remains an optimal solution to problem P4 for any arbitrary angle $ \widetilde \vartheta $ (where $ {\text{e}}^{\jmath \widetilde \vartheta} $ denotes a unit-modulus complex number). The reason is that, the phase of $ \mathbf{u}_l^{\mathsf{H}} \tilde{\mathbf{y}}_l $ does not affect the resulting SNR. Thus, we can constrain the term $ {{{\bf{u}}_l^{\mathsf{H}}}({{\bf{I}}_{K + 1}} \otimes {{\bf{G}}_l})\widetilde {\bf{w}}} $ to be a non-negative real value. As a result, the radar SNR constraint specified in \eqref{34e} can be reformulated as
\begin{align}
	\Re \left\{{{\bf{u}}_l^{\mathsf{H}}}({{\bf{I}}_{K + 1}} \otimes {{\bf{G}}_l})\widetilde {\bf{w}}\right\} \ge {\Delta _{3,l}},l \in {\cal L},\label{48}
\end{align}
where $ {\Delta _{3,l}} \buildrel \Delta \over = \sqrt {{{{\Gamma _l}\varepsilon _l^2{{\bf{u}}_l^{\mathsf{H}}}{\bf{u}}_l} \mathord{\left/
			{\vphantom {{{\Gamma _l}\varepsilon _l^2{{\bf{u}}^{\mathsf{H}}}{\bf{u}}} {(Q\sigma _l^2)}}} \right.
			\kern-\nulldelimiterspace} {(Q\sigma _l^2)}}} $.
Therefore, with the obtained $ {{\bf{w}}_k} $, $ {{\bf{w}}_\vartheta} $, and $ {\bf{u}}_l $, the passive beamforming optimization problem can be reformulated as
\begin{subequations}
	\begin{align}
		{\rm{P5}}:\mathop {\max }\limits_{{\bf{\Phi }},{\eta _k},{\tau _k},{\zeta _1}} \; &\sum\limits_{k = 2}^K {({\eta _k} - {\tau _k})}  + {\zeta _1},\\
		\mathrm{s.t.}\quad\ &\eqref{24c},\eqref{24d},\eqref{24g},\eqref{26}\text{-}\eqref{28},\eqref{48}.
	\end{align}
\end{subequations}
Define $ {{\bf{v}}^{\mathsf{H}}} = [{v_1},{v_2},\ldots,{v_N}] $ where $ {v_n} = {{\text{e}}^{{\text{j}}{\theta _n}}}, n \in {\cal N} $. Then, we can obtain
\begin{align}
	\left({\bf{h}}_{d,i}^{\mathsf{H}} + {\bf{h}}_{r,i}^{\mathsf{H}}{\bf{\Phi G}}\right){{\bf{w}}_j} &= [{\bf{h}}_{r,i}^{\mathsf{H}} {\mathsf{diag}}\{ {\bf{G}}{{\bf{w}}_j}\} \ {\bf{h}}_{d,i}^{\mathsf{H}}{{\bf{w}}_j}]\Big[ \begin{array}{l}
		{\bf{v}}\\ 1
	\end{array} \Big] \notag \\
	&= \widetilde {\bf{H}}_{i,j}^{\mathsf{H}}\widetilde {\bf{v}},i,j \in \{ {\cal K},\vartheta \}. \label{47}
\end{align}
Based on \eqref{47}, the optimization problem P5 can be reformulated as
\begin{subequations}
	\begin{align}
		&{\rm{P6}}:\mathop {\max }\limits_{\widetilde{\bf{v}},{\eta _k},{\tau _k},{\zeta _1}} \;\sum\limits_{k = 2}^K {({\eta _k} - {\tau _k})}  + {\zeta _1},\\
		&{\mathrm{s.t.}}\ \ \sum\limits_{i = 1}^k {{{\left| {\widetilde {\bf{H}}_{k,i}^{\mathsf{H}}\widetilde {\bf{v}}} \right|}^2} + } \sigma _k^2 \ge {{\text{e}}^{{\eta _k}}},k \in {{\cal K}_1},\\
		&\sum\limits_{i = 1}^{k - 1} {{{\left| {\widetilde {\bf{H}}_{k,i}^{\mathsf{H}}\widetilde {\bf{v}}} \right|}^2} + \sigma _k^2}  \le {{\text{e}}^{{{\widehat \tau }_k}}}(1 + {\tau _k} - {\widehat \tau _k}),k \in {{\cal K}_1},\\
		&{\left| {\widetilde {\bf{H}}_{1,1}^{\mathsf{H}}\widetilde {\bf{v}}} \right|^2} + \sigma _1^2 \ge {{\text{e}}^{{\zeta _1}}},\\
		&\Re \left\{{{\bf{u}}_l^{\mathsf{H}}}({{\bf{I}}_{K + 1}} \otimes {{\bf{G}}_l})\widetilde {\bf{w}}\right \} \ge {\Delta _{3,l}},l \in {\cal L},\\
		&{\left| {\widetilde {\bf{H}}_{j,k}^{\mathsf{H}}\widetilde {\bf{v}}} \right|^2} \ge r_{\text{th}}^k\left(\sum\limits_{i = 1}^{k - 1} {{{\left| {\widetilde {\bf{H}}_{j,i}^{\mathsf{H}}\widetilde {\bf{v}}} \right|}^2} + \sigma _j^2} \right),k \in {{\cal K}_1},j = 1,\ldots,k,\\
		&\left\{ \begin{array}{l}
			{\left| {\widetilde {\bf{H}}_{k,\vartheta }^{\mathsf{H}}\widetilde {\bf{v}}} \right|^2} \ge {\left|  {\widetilde {\bf{H}}_{k,K}^{\mathsf{H}}\widetilde {\bf{v}}} \right|^2},k \in {\cal K},\\
			{\left| {\widetilde {\bf{H}}_{k,i + 1}^{\mathsf{H}}\widetilde {\bf{v}}} \right|^2} \ge {\left|  {\widetilde {\bf{H}}_{k,i}^{\mathsf{H}}\widetilde {\bf{v}}} \right|^2},i \in \{ 1,\ldots,K - 1\},
		\end{array} \right.\\
		&\left| {{v_n}} \right| = 1,n \in {\cal N}.
	\end{align}
\end{subequations}
\noindent By incorporating the principle of (\ref{38}) into problem P6, the reformulated optimization problem P7 is presented at the top of this page.
\begin{figure*}[ht]
	\begin{subequations}
		{\small \begin{align}
			{\rm{P7}}:&\mathop {\max }\limits_{\widetilde {\bf{v}},{\eta _k},{\tau _k},{\zeta _1}} \;\sum\limits_{k = 2}^K {({\eta _k} - {\tau _k})}  + {\zeta _1},\\
			{\mathrm{s.t.}}&\ \sum\limits_{i = 1}^k {\left(2\Re \left\{{{\widehat {\widetilde {\bf{v}}}}^{\mathsf{H}}}{{\widetilde {\bf{H}}}_{k,i}}\widetilde {\bf{H}}_{k,i}^{\mathsf{H}}\widetilde {\bf{v}}\right\} - \Re \left\{{{\widehat {\widetilde {\bf{v}}}}^{\mathsf{H}}}{{\widetilde {\bf{H}}}_{k,i}}\widetilde {\bf{H}}_{k,i}^{\mathsf{H}}\widehat {\widetilde {\bf{v}}}\right\}\right)}  + \sigma _k^2 \ge {{\text{e}}^{{\eta _k}}},k \in {{\cal K}_1},\\
			&\left\| {{{\left[2\widetilde {\bf{H}}_{k,k - 1}^{\mathsf{H}}\widetilde {\bf{v}},\ldots,2\widetilde {\bf{H}}_{k,1}^{\mathsf{H}}\widetilde {\bf{v}},2{\sigma _k},{\Delta _1} - 1\right]}^{\mathsf{H}}}} \right\| \le {\Delta _1} + 1,k \in {{\cal K}_1},\\
			&2\Re \left\{{\widehat {\widetilde {\bf{v}}}^{\mathsf{H}}}{\widetilde {\bf{H}}_{1,1}}\widetilde {\bf{H}}_{1,1}^{\mathsf{H}}\widetilde {\bf{v}}\right\} - \Re \left\{{\widehat {\widetilde {\bf{v}}}^{\mathsf{H}}}{\widetilde {\bf{H}}_{1,1}}\widetilde {\bf{H}}_{1,1}^{\mathsf{H}}\widehat {\widetilde {\bf{v}}}\right\} + \sigma _1^2 \ge {{\text{e}}^{{\zeta _1}}},\\
			&\Re \left\{{{\bf{u}}_l^{\mathsf{H}}}({{\bf{I}}_{K + 1}} \otimes {{\bf{G}}_l})\widetilde {\bf{w}}\right\} \ge {\Delta _{3,l}},l \in L, \label{49e}\\
			&2\Re \left\{{\widehat {\widetilde {\bf{v}}}^{\mathsf{H}}}{\widetilde {\bf{H}}_{j,k}}\widetilde {\bf{H}}_{j,k}^{\mathsf{H}}\widetilde {\bf{v}}\right\} - \Re \left\{{\widehat {\widetilde {\bf{v}}}^{\mathsf{H}}}{\widetilde {\bf{H}}_{j,k}}\widetilde {\bf{H}}_{j,k}^{\mathsf{H}}\widehat {\widetilde {\bf{v}}}\right\} \ge r_{\text{th}}^k\left(\sum\limits_{i = 1}^{k - 1} {2\Re \left\{{{\widehat {\widetilde {\bf{v}}}}^{\mathsf{H}}}{{\widetilde {\bf{H}}}_{j,i}}\widetilde {\bf{H}}_{j,i}^{\mathsf{H}}\widetilde {\bf{v}}\right\} - \Re \left\{{{\widehat {\widetilde {\bf{v}}}}^{\mathsf{H}}}{{\widetilde {\bf{H}}}_{j,i}}\widetilde {\bf{H}}_{j,i}^{\mathsf{H}}\widehat {\widetilde {\bf{v}}}\right\}}  + \sigma _j^2\right),k \in {{\cal K}_1},j = 1,\ldots,k,\\
			&\left\{ \begin{array}{l}
				2\Re \left\{{\widehat {\widetilde {\bf{v}}}^{\mathsf{H}}}{\widetilde {\bf{H}}_{k,\vartheta }}\widetilde {\bf{H}}_{k,\vartheta }^{\mathsf{H}}\widetilde {\bf{v}}\right\} - \Re \left\{{\widehat {\widetilde {\bf{v}}}^{\mathsf{H}}}{\widetilde {\bf{H}}_{k,\vartheta }}\widetilde {\bf{H}}_{k,\vartheta }^{\mathsf{H}}\widehat {\widetilde {\bf{v}}}\right\} \ge {\left| {\widetilde {\bf{H}}_{k,K}^{\mathsf{H}}\widetilde {\bf{v}}} \right|^2},k \in {\cal K}\\
				2\Re \left\{{\widehat {\widetilde {\bf{v}}}^{\mathsf{H}}}{\widetilde {\bf{H}}_{k,i + 1}}\widetilde {\bf{H}}_{k,i + 1}^{\mathsf{H}}\widetilde {\bf{v}}\right\} - \Re \left\{{\widehat {\widetilde {\bf{v}}}^{\mathsf{H}}}{\widetilde {\bf{H}}_{k,i + 1}}\widetilde {\bf{H}}_{k,i + 1}^{\mathsf{H}}\widehat {\widetilde {\bf{v}}}\right\} \ge {\left| {\widetilde {\bf{H}}_{k,i}^{\mathsf{H}}\widetilde {\bf{v}}} \right|^2},i \in \{ 1,\ldots,K - 1\} 
			\end{array} \right.,\\
			&\left| {{v_n}} \right| = 1,n \in {\cal N}, \label{49h}
		\end{align}}
	\end{subequations}
	{\noindent}	\rule[10pt]{18cm}{0.04em}
\end{figure*}%
For constraint \eqref{49e}, we first utilize the transformations $ {\bf{\Phi }}{{\bf{g}}_{r,l}} \buildrel \Delta \over = {\mathsf{diag}}\{ {{\bf{g}}_{r,l}}\} {\bf{v}} $ and $ {\mathsf{vec}}\{ ABC\} = ({C^{\mathsf{H}}} \otimes A){\mathsf{vec}}\{ B\} $ to rewrite the term $ ({{\bf{I}}_{K + 1}} \otimes {{\bf{G}}_l})\widetilde {\bf{w}} $ into \eqref{50}, which is shown at the bottom of this page. Thus, the constraint \eqref{49e} can be further rearranged as
\begin{figure*}[hb]
	{\noindent}	\rule[-10pt]{18cm}{0.04em}
	{\small \begin{align}
		&({{\bf{I}}_{K + 1}} \otimes {{\bf{G}}_l})\widetilde {\bf{w}} \notag \\
		&= \left({{\bf{I}}_{K + 1}} \otimes {{\bf{g}}_{d,l}}{\bf{g}}_{d,l}^{\mathsf{H}}\right)\widetilde {\bf{w}} + {\mathsf{vec}}\left\{ {{\bf{G}}^{\mathsf{H}}}{\mathsf{diag}}\{ {{\bf{g}}_{r,l}}\} {\bf{vg}}_{d,l}^{\mathsf{H}}{\bf{W}} + {{\bf{g}}_{d,l}}{{\bf{v}}^{\mathsf{H}}}{\mathsf{diag}}\{ {{\bf{g}}_{r,l}}\} {\bf{GW}} + {{\bf{G}}^{\mathsf{H}}}{\mathsf{diag}}\{ {{\bf{g}}_{r,l}}\} {\bf{v}}{{\bf{v}}^{\mathsf{H}}}{\mathsf{diag}}\{ {{\bf{g}}_{r,l}}\} {\bf{GW}}\right\} \notag \\
		&= \left({{\bf{I}}_{K + 1}} \otimes {{\bf{g}}_{d,l}}{\bf{g}}_{d,l}^{\mathsf{H}}\right)\widetilde {\bf{w}} + \underbrace {\left({{\bf{W}}^{\mathsf{H}}}{{\bf{g}}_{d,l}} \otimes {{\bf{G}}^{\mathsf{H}}}{\mathsf{diag}}\{ {{\bf{g}}_{r,l}}\}  + {{\bf{W}}^{\mathsf{H}}}{{\bf{G}}^{\mathsf{H}}}{\mathsf{diag}}\{ {{\bf{g}}_{r,l}}\}  \otimes {{\bf{g}}_{d,l}}\right)}_{{\bf{F}}_l}{\bf{v}} \notag \\
		&\quad\ + \underbrace {\left({{\bf{W}}^{\mathsf{H}}}{{\bf{G}}^{\mathsf{H}}}{\mathsf{diag}}\{ {{\bf{g}}_{r,l}}\}  \otimes {{\bf{G}}^{\mathsf{H}}}{\mathsf{diag}}\{ {{\bf{g}}_{r,l}}\} \right)}_{{\bf{L}}_l} {\mathsf{vec}}\{ {\bf{v}}{{\bf{v}}^{\mathsf{H}}}\}, \label{50}
	\end{align}}
\end{figure*}%
\begin{align}
	&\Re \left\{{{\bf{u}}_l^{\mathsf{H}}}\left({{\bf{I}}_{K + 1}} \otimes {{\bf{g}}_{d,l}}{\bf{g}}_{d,l}^{\mathsf{H}}\right)\widetilde {\bf{w}} + {{\bf{u}}_l^{\mathsf{H}}}{\bf{F}}_l{\bf{v}} + {{\bf{u}}_l^{\mathsf{H}}}{\bf{L}}_l {\mathsf{vec}}\left\{ {\bf{v}}{{\bf{v}}^{\mathsf{H}}}\right\} \right\} \notag \\
	&= \Re \left\{{{\bf{u}}_l^{\mathsf{H}}}\left({{\bf{I}}_{K + 1}} \otimes {{\bf{g}}_{d,l}}{\bf{g}}_{d,l}^{\mathsf{H}}\right)\widetilde {\bf{w}} + {{\bf{u}}_l^{\mathsf{H}}}{\bf{F}}_l{\bf{v}} + {{\bf{v}}^{\mathsf{H}}}\widetilde {\bf{L}}_l{\bf{v}}\right\} \ge {\Delta _{3,l}},\notag \\
    &\qquad\qquad\qquad\qquad\qquad\qquad\qquad\qquad\qquad\qquad l \in {\cal L},\label{51}
\end{align}
where we derive the matrix $ \widetilde {\bf{L}}_l \in {\mathbb{C}^{N \times N}} $ by reshaping $ {{\bf{L}}_l^{\mathsf{H}}}{{\bf{u}}_l^*} $. Note that the third term in \eqref{51} introduces non-concavity. To address this challenge, we perform a real-valued transformation by defining $ \overline {\bf{v}} \buildrel \Delta \over = {\left[\Re \{{{\bf{v}}^{\mathsf{H}}}\} \Im \{{{\bf{v}}^{\mathsf{H}}}\}\right]^{\mathsf{H}}} $ and $ \overline {\bf{L}}_l \buildrel \Delta \over = \left[ \begin{array}{l}
	- \Re \{\widetilde {\bf{L}}_l\} \ \Im \{\widetilde {\bf{L}}_l\}\\
	\ \ \Im \{\widetilde {\bf{L}}_l\} \ \Re \{\widetilde {\bf{L}}_l\}
\end{array} \right] $. Through this transformation, the quadratic form $ \Re \left\{{{\bf{v}}^{\mathsf{H}}}\widetilde {\bf{L}}_l{\bf{v}}\right\} $ can be equivalently rewritten as $ - {\overline {\bf{v}} ^{\mathsf{H}}}\overline {\bf{L}}_l \overline {\bf{v}} $. Since $ \overline {\bf{L}}_l $ is an indefinite matrix, $ {\overline {\bf{v}} ^{\mathsf{H}}}\overline {\bf{L}}_l \overline {\bf{v}} $ exhibits mixed curvatures across the optimization space. This curvature conflict renders first-order Taylor approximations (e.g., as in \eqref{38}) unreliable for bounding the function.
Therefore, given the solution $ \widehat {\bf{v}} $ obtained in the previous iteration, we construct an approximate upper-bound for $ {\overline {\bf{v}} ^{\mathsf{H}}}\overline {\bf{L}}_l \overline {\bf{v}} $ using the second-order Taylor approximation\footnote{The second-order Taylor approximation retains the curvature information through the quadratic term, enabling a more accurate local approximation of $ {\overline {\bf{v}} ^{\mathsf{H}}}\overline {\bf{L}}_l \overline {\bf{v}} $ and ensuring that the optimization process converges to a bounded solution.} as
\begin{align}
	&{\overline {\bf{v}} ^{\mathsf{H}}}\overline {\bf{L}}_l \overline {\bf{v}} \notag \\
	&\le {\widehat {\bf{v}}^{\mathsf{H}}}\overline {\bf{L}}_l \widehat {\bf{v}} + {\widehat {\bf{v}}^{\mathsf{H}}}\left(\overline {\bf{L}}_l  + {\overline {\bf{L}}_l ^{\mathsf{H}}}\right)\left(\overline {\bf{v}} - \widehat {\bf{v}}\right) + \frac{\lambda }{2}{\left(\overline {\bf{v}}  - \widehat {\bf{v}}\right)^{\mathsf{H}}}\left(\overline {\bf{v}}  - \widehat {\bf{v}}\right) \notag \\
	&= \Re \left\{{\widehat {\bf{v}}^{\mathsf{H}}}\left(\overline {\bf{L}}_l  + {\overline {\bf{L}}_l ^{\mathsf{H}}} - \lambda {I_{2N}}\right){\bf{Uv}}\right\} - {\widehat {\bf{v}}^{\mathsf{H}}}{\overline {\bf{L}}_l ^{\mathsf{H}}}\widehat {\bf{v}} + \lambda N, \label{52}
\end{align}
where $ \lambda $ denotes the maximum eigenvalue of matrix $ \left(\overline {\bf{L}}_l  + {\overline {\bf{L}}_l ^{\mathsf{H}}}\right) $, and $ {\bf{U}} \buildrel \Delta \over = {[{\bf{I}}_{N \times N} \ {\text{j}}{\bf{I}}_{N \times N}]^{\mathsf{H}}} $ is utilized to transform a real-valued expression into a complex-valued one. Leveraging the unit-modulus property of the reflecting coefficients, we know that $ {\overline {\bf{v}} ^{\mathsf{H}}}\overline {\bf{v}}  = {\widehat {\bf{v}}^{\mathsf{H}}}\widehat {\bf{v}} = N $. Substituting the approximation \eqref{52} into \eqref{51}, we can reformulate the radar SNR constraint for each iteration as follows:
\begin{align}
	\Re \left\{{\widetilde {\bf{u}}_l^{\mathsf{H}}}{\bf{v}}\right\} \le {\Delta _{4,l}},l \in {\cal L},
\end{align}
where
\begin{equation}
\begin{split}
    {\Delta _{4,l}} \buildrel \Delta \over =&  - {\Delta _{3,l}} + {\widehat {\bf{v}}^{\mathsf{H}}}{\overline {\bf{L}}_l ^{\mathsf{H}}}\widehat {\bf{v}} \\&+ \Re \left \{{{\bf{u}}_l^{\mathsf{H}}}\left({{\bf{I}}_{K + 1}} \otimes {{\bf{g}}_{d,l}}{\bf{g}}_{d,l}^{\mathsf{H}}\right)\widetilde {\bf{w}} \right \} - \lambda N,
\end{split}
\end{equation}
and $ \widetilde {\bf{u}}_l \buildrel \Delta \over = {\left( - {{\bf{u}}_l^{\mathsf{H}}}{\bf{F}} + {\widehat {\bf{v}}^{\mathsf{H}}}\left(\overline {\bf{L}}_l  + {\overline {\bf{L}}_l ^{\mathsf{H}}} - \lambda {I_{2N}}\right){\bf{U}}\right)^{\mathsf{H}}} $.

In problem P7, the remaining challenge lies in dealing with the unit-modulus constraint \eqref{49h}, which can be addressed through the penalty convex-concave procedure \cite{10776025, 10472878}. Specifically, the unit-modulus constraint can be rewritten as
\begin{align}
	1 \le {\left| {{v_n}} \right|^2} \le 1,n \in {\cal N}.
\end{align}
Based on \eqref{38}, the part $ 1 \le {\left| {{v_n}} \right|^2} $ can be recast as
\begin{align}
	1 \le 2\Re \left\{\widehat v_n^{\mathsf{H}}{v_n}\right\}- \Re \left\{\widehat v_n^{\mathsf{H}}{\widehat v_n}\right\}, n \in {\cal N}.
\end{align}
Following the operations described above, all the constraints in problem P7 are convex. As a consequence, it can be efficiently solved using well-established toolboxes such as CVX.

\begin{algorithm}[t] 
	\caption{AO Algorithm for Solving Problem P1.} 
	\begin{algorithmic}[1]
		\STATE
		\textbf{Initialize:} $ {\bf{\Phi }}^{(0)} $ and $ {\bf{u}}_l^{(0)} $, iteration index $ t = 1 $ and accuracy threshold $ \varepsilon > 0 $.
		\STATE
		\textbf{Repeat:}
		\STATE
		In the $ t $-th iteration, with the given $ {\bf{\Phi }}^{(t-1)} $ and $ {\bf{u}}_l^{(t-1)} $, solve the problem P3 to obtain $ {\mathbf{w}}_k^{(t)} $ and $ {\mathbf{w}}_\vartheta^{(t)} $.
		\STATE
		Update $ {{\bf{u}}_l^{(t)}} = \frac{{({{\bf{I}}_{K + 1}} \otimes {{\bf{G}}_l})\widetilde {\bf{w}}}}{{{{\widetilde {\bf{w}}}^{\mathsf{H}}}({{\bf{I}}_{K + 1}} \otimes {\bf{G}}_l^{\mathsf{H}}{{\bf{G}}_l})\widetilde {\bf{w}}}} $. 
		\STATE
		With obtained $ {\mathbf{w}}_k^{(t)} $, $ {\mathbf{w}}_\vartheta^{(t)} $ and $ {{\bf{u}}_l^{(t)}} $, solve the problem P7 to obtain $ {\bf{\Phi }}^{(t)} $.
		\STATE
		Update $ t = t + 1 $.
		\STATE
		\textbf{Until:} the increase of the objective function between two
		adjacent iterations in P3 is smaller than $ \varepsilon $.
		\label{algorithm1} 
	\end{algorithmic} 
\end{algorithm}

\subsection{Complexity Analysis}

In the preceding subsections, the original problem P1 is first decomposed into three subproblems. Subsequently, the AO and Taylor approximation methods are utilized to solve these subproblems. The steps for solving problem P1 are summarized in Algorithm 1.

According to \cite{11086503, 6891348}, the complexity depends on the number of variables and constraints of the optimization problem. For Algorithm 1, the overall complexity is dominated by the complexity for solving problems P3 and P7. Specifically, the number of iterations for solving P3 is $ {I_1} = \sqrt {K(3K + 7)/2 + L-1} $, and the complexity of each iteration is $ {{\bf{\psi }}_1} = {l_1}(3K(K + 1)/2 + L - 1 + (K - 1){(K + 1)^2} + {(MK + M)^2}) + l_1^2(3K(K + 1)/2 + L - 1) + l_1^3 $, where $ {l_1} = 2(M + K) - 1 $. Therefore, the complexity for sovling problem P3 is $ {\cal O}({I_1}{n_1}) $. Similarly, the complexity of problem P7 can be expressed as $ {\cal O}({I_2}{n_2}) $, where $ {I_2} = \sqrt {K(3K + 7)/2 + 2N + L - 3} $ denotes the iteration number, and $ {n_2} = {l_2}(3K(K + 1)/2 + L - 1 + (K - 1){(K + 1)^2} + 4N) + l_2^2(3K(K + 1)/2 + L - 1) + l_2^3 $, where $ {l_2} = N + 2K $ represents the complexity of each iteration.

Finally, denoting $ I_{\text{AO}} $ as the number of iterations of the AO algorithm (i.e., Algorithm 1), the overall complexity of Algorithm 1 is $ {\cal O}({I_{\text{AO}}}({I_1}{n_1} + {I_2}{n_2})) $.

\section{Numerical Results}
In this section, simulation results are presented to evaluate the performance of the proposed RIS-empowered NOMA-ISAC system. We set $ M = 6 $, $ K = 4 $, $ L = 2 $, $ \varepsilon _l^2 = \sigma _k^2 = - 90{\mathrm{dBm}}, \sigma _l^2 = 1, \forall l,\forall k $, and $ Q = 1024 $. The distances of the DFBS-RIS, RIS-target, and RIS-user links are set to be $ 40{\mathrm{m}} $, $ 4{\mathrm{m}} $, and $ 8{\mathrm{m}} $ respectively. A typical distance-dependent path-loss model, as described in \cite{8811733}, is adopted. The path-loss exponents for the DFBS-RIS, RIS-target, RIS-user, DFBS-target, and DFBS-user links are set as $ 1.1 $, $ 1.1 $, $ 1.2 $, $ 1.2 $, and $ 1.7 $, respectively. Since users are located several meters farther from the target, reflected signals from the target to users are neglected due to severe channel fading. Additionally, the Rician factor for the DFBS-RIS/user and RIS-user links is set as $ \kappa = 3{\mathrm{dB}} $. 

\subsection{Illustration of Radar Sensing Performance}
\begin{figure}[t]
	\centering
	\includegraphics[width=0.92\linewidth,height=0.66\linewidth]{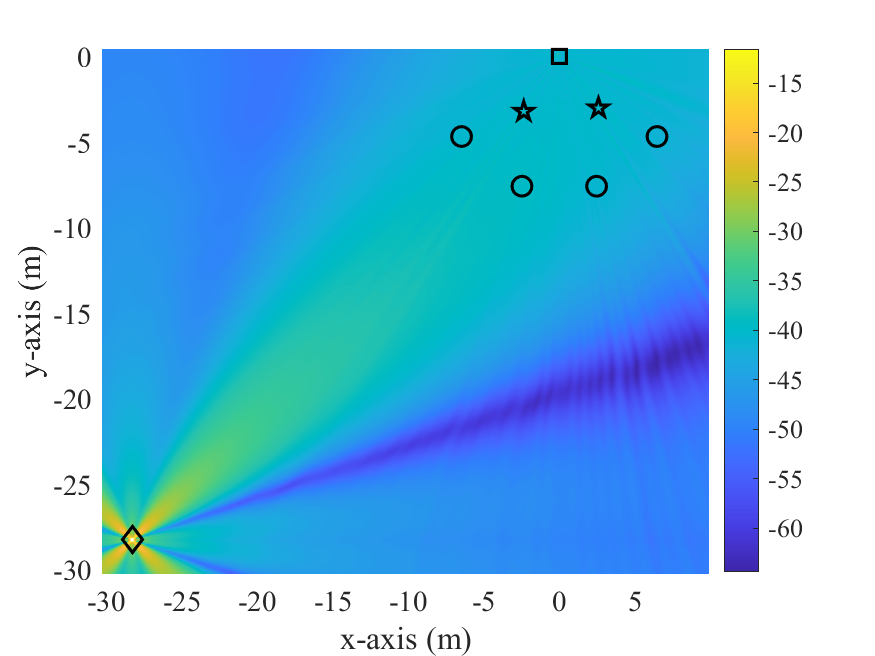}
	\caption{Enhanced beampattern of the RIS-empowered NOMA-ISAC system (DFBS: diamond; RIS: square; targets: stars; users: circles).}
	\label{fig:3}
\end{figure}
\begin{figure}[t]
	\centering
	\includegraphics[width=0.92\linewidth,height=0.66\linewidth]{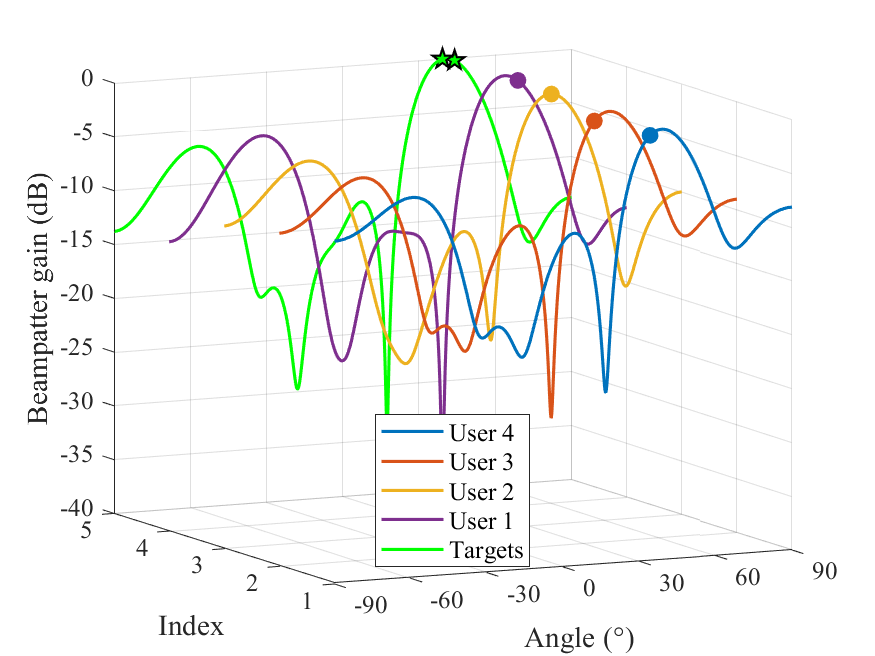}
	\caption{The DFBS transmit beampattern versus angle (targets: stars; users: solid circles).}
	\label{fig:4}
\end{figure}

The beampattern gain characterizes the spatial distribution of signal energy radiated by the joint DFBS-RIS system. For any spatial point \((x,y)\), it is defined as \cite{10364735}
\begin{align}
    BP(x,y) = \left\| \left( \mathbf{h}_{d}^{{\mathsf{H}}}(x,y) + \mathbf{h}_{r}^{{\mathsf{H}}}(x,y) \boldsymbol{\Phi} \mathbf{G} \right) \mathbf{W} \right\|^{2},
\end{align} 
where $ \mathbf{h}_{d}(x,y) $ and $ \mathbf{h}_{r}(x,y) $ are the channel vectors from the DFBS and RIS to the point $ (x,y) $, respectively. In Fig. \ref{fig:3}, we provide a 2D visualization of the signal energy distribution. 
It can be observed that the DFBS generates strong beams toward the locations of the RIS, targets, and users. Meanwhile, the RIS further forms multiple passive beams to guide signals toward these regions.

The transmit beampattern of the DFBS is shown in Fig. \ref{fig:4}. The beampattern emitted by the DFBS accurately focuses the main lobe on the direction of the targets and users. Meanwhile, the sidelobe level in other directions is very low. This verifies that the proposed RIS-empowered NOMA-ISAC scheme can efficiently assist communication and sensing functions. It is also consistent with the results in Fig. \ref{fig:3}.

\subsection{Illustration of Communication Sum-rate Performance}
To validate the advantages of the proposed algorithm, several baseline algorithms are introduced as follows:
\begin{itemize}
	\item ``Comm only'': In this case, for the considered system, only the multi-user communication function is optimized, with radar sensing constraints excluded.
	
	\item ``Discrete phase'': This algorithm assumes that the phase shift of each RIS element can only take a finite set of discrete values.
	
	\item ``Random phase'': In this algorithm, only transmit beamforming and receive filtering are optimized, with the RIS adopting random phase shifts.
	
	\item ``Without RIS'': This algorithm eliminates the RIS-aided communication and sensing pathways, operating as a direct NOMA-ISAC system without RIS assistance.
	
	\item ``Without NOMA'': NOMA technology is dispensed with in this setup, with the focus shifted to a RIS-empowered ISAC system that uses traditional multiple access techniques for signal transmission.
\end{itemize}

\begin{figure}[t]
	\centering	\includegraphics[width=0.92\linewidth,height=0.66\linewidth]{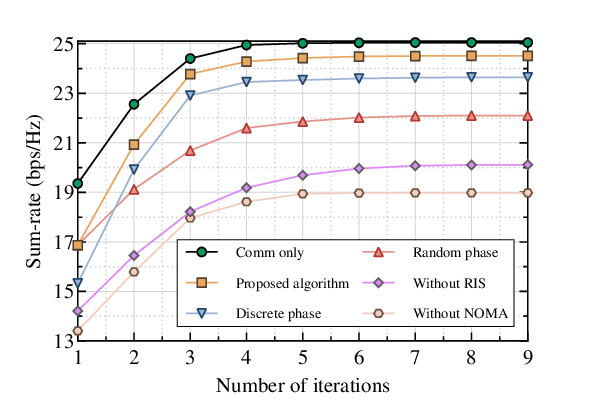}
	\caption{The convergence behavior of the proposed algorithm and the
	baseline algorithms.}
	\label{fig:5}
\end{figure}
\begin{figure}[t]
	\centering
	\includegraphics[width=0.92\linewidth,height=0.66\linewidth]{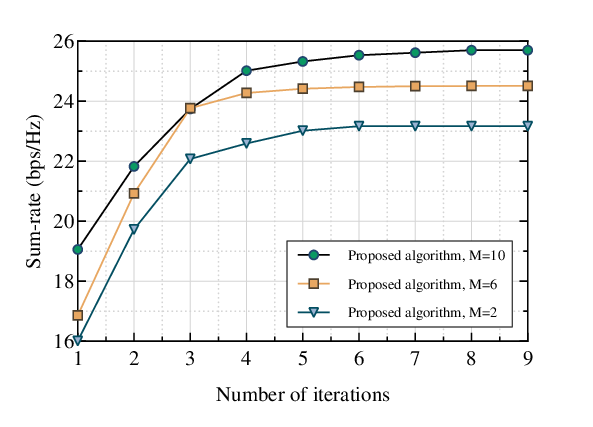}
	\caption{The convergence behavior of the proposed algorithm with different $ M $.}
	\label{fig:6}
\end{figure}
Fig. \ref{fig:5} presents a convergence comparison of all algorithms when $ M = 6 $, $ N = 60 $, and $ P_{\text{th}} = 40{\mathrm{dBm}} $. Notably, all algorithms converge within $ 9 $ iterations. The ``comm only" algorithm achieves the optimal sum-rate performance due to the absence of radar constraints. The proposed algorithm outperforms the remaining four baseline algorithms. Specifically, compared with the ``discrete phase" algorithm, the proposed algorithm features higher phase adjustment precision and degrees of freedom, enabling superior signal manipulation and system performance. Against the ``random phase" algorithm, its advantage stems from the optimized phase shifts of the RIS. For the ``without RIS" algorithm, the performance improvement is attributed to the RIS integration, which creates additional communication links to enhance channel gains. Another key finding is that the algorithms incorporating NOMA technology outperform without NOMA algorithm, as NOMA allows multiple users to multiplex transmissions over the same time-frequency resources, significantly improving spectral efficiency and system sum-rate.

The sum-rate of the ``proposed algorithm'' corresponding to different values of $ M $ is presented in Fig. \ref{fig:6} when $ N = 60 $ and $ P_{\text{th}} = 40{\mathrm{dBm}} $. Evidently, the increase in the number of DFBS antennas results in a promotion of the system's sum-rate. The underlying reason lies in the fact that with the growth of $ M $, the beamforming gain and spatial resource experience a corresponding increase, which in turn augments the transmission rate. 

\begin{figure}[t]
	\centering
	\includegraphics[width=0.92\linewidth,height=0.66\linewidth]{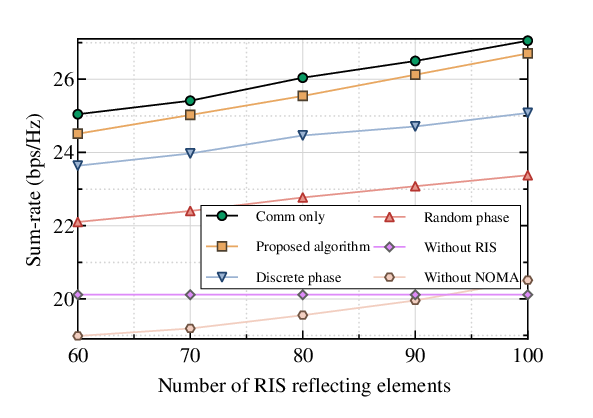}
	\caption{The sum-rate performance versus the number of RIS reflecting elements.}
	\label{fig:7}
\end{figure}
\begin{figure}[t]
	\centering
	\includegraphics[width=0.92\linewidth,height=0.66\linewidth]{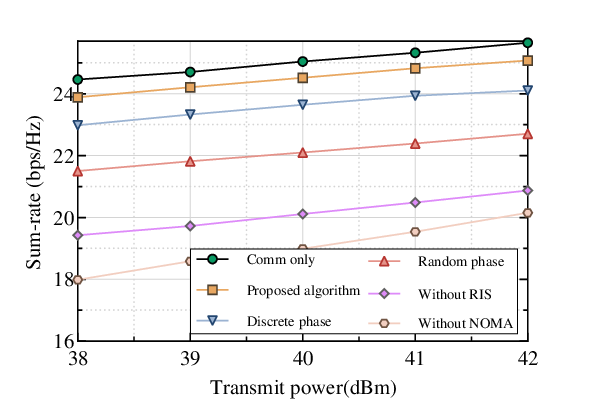}
	\caption{The sum-rate performance versus the transmit power of the DFBS.}
	\label{fig:8}
\end{figure}

Fig. \ref{fig:7} illustrates the sum-rate performance variation with the number of reflecting elements $ N $ under $ M = 6 $ and $ P_{\text{th}} = 40{\mathrm{dBm}} $. For all algorithms except ``without RIS", the sum-rate gradually increases as $ N $ grows from 60 to 100. This trend stems from the fact that more reflecting elements enable stronger channel gains: each additional element contributes to cumulative signal reflection, which enhances the constructive interference of transmitted signals. The resulting increase in received signal power directly improves the system sum-rate, demonstrating the critical role of RIS element quantity in boosting communication efficiency.

In Fig. \(\ref{fig:8}\), when \(M = 6\) and \(N = 60\), the relationship between the sum-rate and the total transmit power \(P_{\text{th}}\) is presented. For all algorithms, it can be observed that the sum-rate increases as \(P_{\text{th}}\) increases. The fundamental reason for this trend lies in the fact that a higher \(P_{\text{th}}\) directly leads to a significant enhancement of the signal power received at users, which in turn improves the SINR and thereby increases the sum-rate.

\begin{figure}[t]
	\centering
    \includegraphics[width=0.92\linewidth,height=0.66\linewidth]{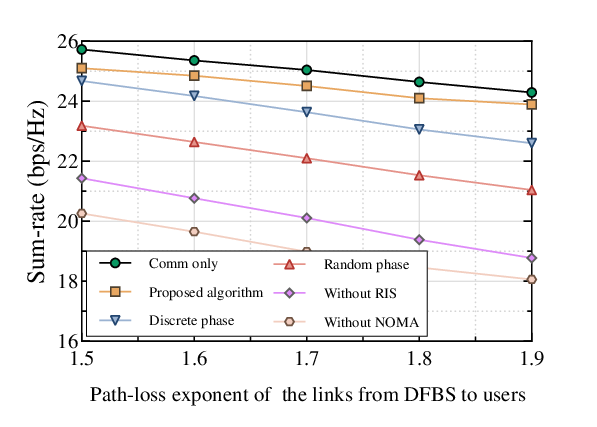}
	\caption{The sum-rate performance versus the path-loss exponent of DFBS-user links.}
	\label{fig:9}
\end{figure}
\begin{figure}[t]
	\centering
	\includegraphics[width=0.92\linewidth,height=0.66\linewidth]{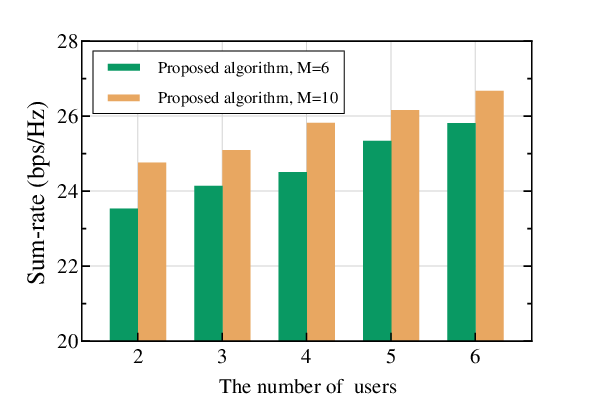}
	\caption{The sum-rate performance versus the number of users.}
	\label{fig:10}
\end{figure}

Fig. \ref{fig:9} shows the effect of the DFBS-to-user path-loss exponent on the sum-rate with $ M = 6 $, $ N = 60 $, and $ P_{\text{th}} = 40{\mathrm{dBm}} $. As the exponent increases, the sum-rate of all algorithms decreases due to signal attenuation, but the ``without RIS" algorithm degrades more sharply, highlighting its vulnerability to severe path loss. In contrast, the RIS-empowered system mitigates this through reconfigurable metasurfaces that create alternative signal paths, compensating for direct-link degradation and maintaining more stable performance. Under high path-loss conditions (e.g., significant direct-link blockage), the RIS system’s resilience is evident, as indirect links counterbalance attenuation to sustain reliable transmission. These results validate RIS’s role in enhancing wireless robustness.

Fig. \ref{fig:10} shows a significant positive correlation between the number of users and the sum-rate with $ N = 60 $ and $ P_{\text{th}} = 40{\mathrm{dBm}} $. The core reason for this trend lies in the enhancement of spectrum reuse efficiency by multi-user resource allocation. When more users access, the system can achieve more efficient data transmission within the same frequency band, thus increasing the resource utilization density per unit spectrum. In addition, this result verifies that the algorithm proposed in this paper can still maintain the high efficiency of spectrum utilization by optimizing resource allocation even in scenarios where the access demand is continuously increasing.

\section{Conclusion}
This paper investigates a downlink NOMA-based ISAC system integrated with RIS. The objective is to maximize the sum-rate of users by jointly optimizing the transmit beamforming vectors, receive filters, and RIS phase-shift vectors. The optimization problem is subject to multiple constraints, including the sensing SNR requirements, SIC decoding conditions, user SINR demands, characteristics of receive filters, power limitations, and properties of RIS reflection coefficients. An efficient algorithm based on the AO and SCA methods is proposed to iteratively solve it. Simulation results demonstrate that the proposed algorithm successfully enhances the sum-rate of communicating users while fulfilling the sensing prerequisites of the targets.

\bibliographystyle{IEEEtran}
\bibliography{reference}

\end{document}